\documentclass{aa}
\usepackage[latin1]{inputenc}
\pdfoutput=1
\usepackage{fancyhdr}
\usepackage{amsmath}
\usepackage{amssymb}
\usepackage{txfonts}
\usepackage{multirow}
\usepackage{longtable}
\usepackage{color}

\usepackage{graphicx}
\usepackage{epsfig}
\usepackage{url}
\usepackage{xspace}

\usepackage[round]{natbib}
\bibpunct{(}{)}{;}{a}{}{,}
\usepackage{aas_macros} 

\begin{document}

\title{The \textit{400d} Galaxy Cluster Survey weak lensing programme: III:\\
Evidence for consistent WL and X-ray masses at $z\!\approx\!0.5$.}
\titlerunning{The \textit{400d} weak lensing survey III}

\author{Holger Israel\inst{1,2,}\thanks{e-mail: holger.israel@durham.ac.uk} 
   \and Thomas H. Reiprich\inst{2}
   \and Thomas Erben\inst{2} 
   \and \\Richard J. Massey\inst{1}
   \and Craig L. Sarazin\inst{3}
   \and Peter Schneider\inst{2}
   \and Alexey Vikhlinin\inst{4}
    }
\institute{Department of Physics, Durham University, South Road, Durham DH1 3LE, UK
\and Argelander-Institut f\"ur Astronomie, Auf dem H\"ugel 71, 53121 Bonn, Germany
 \and Department of Astronomy, University of Virginia, 530 McCormick Road, Charlottesville, VA 22904, USA
 \and Harvard-Smithsonian Center for Astrophysics, 60 Garden Street, Cambridge, MA 02138, USA}

\date{Received / Accepted }

\abstract{
Scaling properties of galaxy cluster observables with cluster mass provide central insights into
the processes shaping clusters.  Calibrating proxies for cluster mass that are 
relatively cheap to observe  will moreover be
crucial to harvest the cosmological information available from the number and growth
of clusters  with upcoming surveys like \textit{eROSITA} and \textit{Euclid}.
The recent \textit{Planck} results led to suggestions that X-ray masses might be 
biased low by $\sim\!40$~\%, more than previously considered.
}{
We aim to extend  knowledge of the weak lensing -- X-ray
mass scaling towards lower masses (as low as $1\!\times\!10^{14}\,\mathrm{M}_{\odot}$) in a sample  
representative of the $z\!\sim\!0.4$--$0.5$ population. Thus, we extend the direct calibration 
of cluster mass estimates to higher redshifts. 
}{
We investigate the scaling behaviour of MMT/Megacam weak lensing (WL) masses for $8$ clusters at 
$0.39\!\leq\!z\!\leq\!0.80$ as part of the \emph{400d} WL programme with hydrostatic 
\textit{Chandra} X-ray masses as well as those based on the proxies, e.g. 
$Y_{\mathrm{X}}\!=\!T_{\mathrm{X}}M_{\mathrm{gas}}$.
}{
Overall, we find good agreement between WL and X-ray masses,  
with different mass bias estimators all consistent with zero.
When subdividing the sample into a low-mass and a high-mass
subsample, we find the high-mass subsample to show no significant
mass bias while for the low-mass subsample, there is a bias towards
overestimated X-ray masses at the $\sim\!2\sigma$ level for some mass proxies.
The overall scatter in the mass-mass scaling relations is surprisingly low.
Investigating possible causes, we find that
neither the greater range in WL than in X-ray masses nor the small scatter can be
traced back to the parameter settings in the WL analysis.
}{
We do not find evidence for a strong ($\sim\!40$~\%)
underestimate in the X-ray masses, as suggested to reconcile recent \textit{Planck} cluster counts and
cosmological constraints.
For high-mass clusters, our measurements are consistent with other studies in the literature.
The mass dependent bias, significant at $\sim\!2\sigma$, may hint at a physically
different cluster population (less relaxed clusters with more substructure and mergers); or it
may be due to small number statistics.  Further 
studies of low-mass high-$z$ lensing clusters will elucidate their mass scaling behaviour.
}


\maketitle

\section{Introduction}

Galaxy cluster masses hold a crucial role in cosmology.
 In the paradigm of hierarchical structure formation from 
tiny fluctuations in the highly homogeneous early cosmos after inflation, clusters emerge via
the continuous matter accretion onto local minima of the gravitational potential.
Depending sensitively on cosmological parameters, the cluster mass function, i.e.\ their 
abundance as function of mass and redshift $z$, provides observational constraints to cosmology
\citep[e.g.,][]{2009ApJ...692.1060V,2011ARA&A..49..409A,2013arXiv1303.5080P}.

Observers use  several avenues to determine cluster masses: properties of the
X-ray--emitting intracluster medium (ICM), its imprint on the cosmic microwave background via
the Sunyaev-Zel'dovich (SZ) effect in the sub-mm regime, galaxy richness estimates and 
dynamical masses via optical imaging and spectroscopy, and gravitational lensing.   
Across all wavelengths, cluster cosmology surveys are under preparation, aiming at a complete 
cluster census  out to ever higher redshifts, e.g. \textit{eROSITA} 
\citep{2010SPIE.7732E..23P,2012arXiv1209.3114M,2012MNRAS.422...44P} and 
\textit{Athena} \citep{2013arXiv1306.2307N,2013arXiv1306.2319P}
in X-rays, \textit{Euclid} \citep{2011arXiv1110.3193L,2012arXiv1206.1225A}, DES and LSST in the 
optical/near-infrared, CCAT \citep{2012SPIE.8444E..2MW} and SKA at sub-mm and radio frequencies.

Careful X-ray studies of  clusters at  low and intermediate redshifts  yield highly precise cluster
masses, but assume hydrostatic equilibrium, and in most cases spherical symmetry 
\citep[e.g.][]{2008A&A...487..431C,2013SSRv..177..119E}. Observational evidence and numerical 
modelling challenge these assumptions for all but the most relaxed systems 
\citep[e.g.][]{2008MNRAS.384.1567M,2012NJPh...14e5018R,2013SSRv..177..155L,2013ApJ...765...24N}.
While  simulations find X-ray masses to only slightly underestimate the true mass
of clusters that exhibit no indications of recent mergers and can be considered virialised,
 non-thermal pressure support can lead to a $>\!20$~\% bias in unrelaxed clusters
\citep{2010A&A...510A..76L,2012NJPh...14e5018R}.
\citet{2014ArXiv1401.7657S} modelled the pressure due to ICM turbulence analytically
and found a $\sim\!10$~\% underestimate of cluster masses compared to the hydrostatic case.

Weak lensing (WL), in contrast, is subject to larger stochastic uncertainties, but can in 
principle yield unbiased masses, as no equilibrium assumptions are required. 
Details of the mass modelling however can introduce biases,
in particular concerning projection effects, the source redshift distribution
 and the departures from an axisymmetric mass profile 
\citep{2009MNRAS.396..315C,2011ApJ...740...25B,2012MNRAS.421.1073B,2013SSRv..177...75H}. 
For individual clusters, stochastic uncertainties dominate the budget; however, larger cluster
samples benefit from improved corrections for  lensing systematics, driven by cosmic shear 
projects \citep[e.g.][]{2013MNRAS.429..661M}.
    
Most of the leverage on cosmology and structure formation from future cluster surveys will be
due to clusters at higher $z$ than have been previously investigated. Hence, 
the average cluster masses and signal/noise ratios for all observables are going to be smaller. 
Even and especially for the deepest surveys, most objects will lie close to the detection limit.
Thus the scaling of inexpensive proxies (e.g. X-ray luminosity $L_{\mathrm{X}}$) with total mass
needs to be calibrated against representative cluster samples at low and high $z$. 
Weak lensing and SZ mass estimates are both good candidates as they exhibit independent 
systematics from X-rays and a weaker $z$-dependence in their signal/noise ratios.

Theoretically, cluster scaling relations arise from their description as self-similar objects
forming through gravitational collapse \citep{1986MNRAS.222..323K}, and deviations from simple
scaling laws provide crucial insights into cluster physics.
For the current state of scaling relation science, we point to the recent review by 
\citet{2013SSRv..177..247G}. As we are interested in the cluster population to be seen by 
upcoming surveys, we focus here on results obtained at high redshifts.

Self-similar modelling includes evolution of the scaling relation \emph{normalisations} with 
the Hubble expansion, which is routinely measured \citep[e.g.][]{2011A&A...535A...4R,
2013AN....334..354E}. Evolution effects beyond self-similarity, e.g.\ due to declining AGN 
feedback at low $z$, have been claimed and discussed
\citep[e.g.][]{2007MNRAS.382.1289P,2010MNRAS.408.2213S,2010ApJ...715.1508S,2012MNRAS.421.1583M},
but current observations are insufficient to constrain possible evolution in
slopes \citep{2013SSRv..177..247G}. 
Evidence for different scaling behaviour in groups and low-mass clusters 
was found by, e.g.,\ \citet{2011A&A...535A.105E,2012MNRAS.422.2213S,2014arXiv1402.0868B}.

\citet{2011A&A...535A...4R} and \citet{2012MNRAS.421.1583M} investigated X-ray scaling 
relationships including clusters at $z\!>\!1$, and both stressed 
the increasing influence of selection effects at higher $z$. 
Larger weak lensing samples of distant clusters are just in the process of being compiled 
\citep{2011ApJ...737...59J,2012A&A...546A.106F,2011ApJ...726...48H,2012MNRAS.427.1298H,2012A&A...546A..79I,2012arXiv1208.0597V,2012ApJS..199...25P}.
Thus most WL scaling studies are currently limited to $z\!\!\lesssim\!\!0.6$, 
and also include nearby 
clusters \citep[e.g.][M13]{2012MNRAS.427.1298H,2013ApJ...767..116M}. The latter authors find 
projected WL masses follow the expected correlation with the  SZ signal $Y_{\mathrm{SZ}}$, corroborating
similar results for more local clusters by \citet{2009ApJ...701L.114M,2012ApJ...754..119M}.
\citet{2013MNRAS.429.3627M} performed a detailed WL analysis of a 
$z\!=\!0.81$ cluster discovered in the SZ using the Atacama Cosmology Telescope, and compared
the resulting lensing mass against the \citet{2012ApJ...751...12R} $Y_{\mathrm{SZ}}$--$M$ 
scaling relation, in what they describe as a first step towards a high-$z$ SZ-WL scaling study.

By compiling \textit{Hubble Space Telescope} data for $27$ massive
clusters at $0.83\!<\!z\!<\!1.46$, \citet{2011ApJ...737...59J} not only derive the 
relation between WL masses $M^{\mathrm{wl}}$ and ICM temperature $T_{\mathrm{X}}$,
but also notice a good correspondence between WL and hydrostatic X-ray masses
$M^{\mathrm{hyd}}$. As they focus on directly testing cosmology 
with the most massive clusters ,  
these authors however stop short of deriving the WL--X-ray scaling. 
Also using \textit{HST} observations, \citet{2011ApJ...726...48H} investigated the WL mass scaling
of the optical cluster richness (i.e.\ galaxy counts) and $L_{\mathrm{X}}$ of $25$ 
moderate-$L_{\mathrm{X}}$ clusters at $0.3\!<\!z\!<\!0.6$, thus initiating the study of
WL scaling relations off the top of the mass function.

Comparisons between weak lensing and X-ray masses for larger cluster samples were pioneered
by \citet{2008MNRAS.384.1567M} and \citet{2008A&A...482..451Z}, collecting evidence for
the ratio  of weak lensing  to X-ray masses $M^{\mathrm{wl}}/M^{\mathrm{hyd}}\!>\!1$, indicating
non-thermal pressure. \citet{2010ApJ...711.1033Z}, analysing $12$ clusters from the 
\textit{Local Cluster Substructure Survey} (LoCuSS), find
this ratio to depend on the radius .
Likewise, a difference between relaxed and unrelaxed clusters is found 
\citep{2010ApJ...711.1033Z,2013ApJ...767..116M}. \citet{2012NJPh...14e5018R} show that the
gap between X-ray and lensing masses is more pronounced in simulations than in observations,
pointing to either an underestimate of the true mass also by WL masses 
\citep[cf.][]{2012MNRAS.421.1073B} or to simulations overestimating the X-ray mass bias.

The current disagreement between the cosmological constraints derived from \textit{Planck} 
primary cosmic microwave background (CMB) data with \textit{Wilkinson Microwave Anisotropy Probe}
data, supernova data, and cluster data \citep{2013arXiv1303.5080P} may well be alleviated by, 
e.g.\ sliding up a bit along the \textit{Planck} degeneracy curve between the Hubble factor $H_{0}$ 
and the matter density parameter $\Omega_{\mathrm{m}}$. Nevertheless, as stronger cluster mass biases
than currently favoured $(\sim\!40$~\%) have also been invoked as a possible explanation,
it is very important to test the cluster mass calibration with independent methods
out to high $z$, as we do in this work. 

This article aims to test the agreement of the weak lensing and X-ray masses measured by
\citet{2012A&A...546A..79I} for $8$ relatively low-mass clusters at $z\!\gtrsim\!0.4$ 
with scaling relations from the recent literature. 
The \emph{400d} X-ray sample from which our clusters are drawn has been constructed to contain
typical objects at intermediate redshifts, similar in mass and redshift to upcoming surveys.
Hence, it does not include extremely massive low-$z$ clusters. 
We describe the observations and WL and X-ray mass measurements for the $8$ clusters in 
Sect.~\ref{sec:obsdat}, before presenting the central 
scaling relations in Sect.~\ref{sec:res}. 
Possible explanations for the steep slopes our scaling relations exhibit are discussed in 
Sect.~\ref{sec:puzzle}, and we compare to literature results in Sect.~\ref{sec:m13disc}, leading
to the conclusions in Sect.~\ref{sec:conclu}. Throughout this article, 
$E(z)\!=\!H(z)/H_{0}\!=\!\sqrt{\Omega_{\mathrm{m}}(z+1)^{3}+\Omega_{\Lambda}}$ denotes the
self-similar evolution factor (Hubble factor $H(z)$ normalised to its present-day value of
$H_{0}\!=\!72\,\mbox{km}\,\mbox{s}^{-1}\,\mbox{Mpc}^{-1}$), 
computed for a flat universe with matter and Dark Energy densities 
of $\Omega_{\mathrm{m}}\!=\!0.3$ and $\Omega_{\Lambda}\!=\!0.7$ in units of the critical density.

\section{Observations and Data Analysis} \label{sec:obsdat}

\subsection{The \emph{400d} weak lensing survey}

This article builds on the weak lensing analysis for $8$ clusters of galaxies 
\citep[Paper~I and Paper~II hereafter]{2010A&A...520A..58I,2012A&A...546A..79I} selected from 
the \emph{400d} X-ray selected sample of clusters 
\citep[V09a]{2007ApJS..172..561B,2009ApJ...692.1033V}.
From the $\sim\!400\,\mbox{deg}^{2}$ of all suitable \textit{Rosat} PSPC observations,
\citet{2007ApJS..172..561B} compiled a catalogue of serendipitously detected clusters, i.e.\
discarding the intentional targets of the \textit{Rosat} pointings. For a uniquely
complete subsample of $36$
X-ray luminous ($L_{\mathrm{X}}\gtrsim 10^{44}\,\mbox{erg/s}$) high-redshift 
($0.35\!\leq\!z\!\leq\!0.89$) sources, V09a obtained deep \textit{Chandra} data, weighing the
clusters using three different mass proxies (Sect.~\ref{sec:ygt}). Starting from the cluster 
mass function computed by V09a, \citet{2009ApJ...692.1060V} went on to constrain cosmological 
parameters. For brevity, we will refer to the V09a high-$z$ sample as the \emph{400d} sample. 
The \emph{400d} weak lensing survey follows up these clusters in weak lensing, determining
independent WL masses with the ultimate goals of deriving the lensing-based mass function for 
the complete sample and to perform detailed consistency checks. Currently, we have determined
WL masses for $8$ clusters observed in four dedicated MMT/Megacam runs (see Papers~I and II).
Thus, our scaling relation studies are largely limited to this subset of clusters, covering the
sky between $\alpha_{\mathrm{J2000}}\!=\!13^{\mathrm{h}}30^{\mathrm{m}}$--$24^{\mathrm{h}}$ 
with $\delta_{\mathrm{J2000}}\!>\!10^{\circ}$ and 
$\alpha_{\mathrm{J2000}}\!=\!0^{\mathrm{h}}$--$08^{\mathrm{h}}30^{\mathrm{m}}$ with 
$\delta_{\mathrm{J2000}}\!>\!0^{\circ}$.

\subsection{Weak lensing analysis} \label{sec:wla}

\begin{table*}
\caption{Measured properties of the \emph{400d} MMT cluster sample. We quote the properties adopted
from V09a and state the overdensity radii and corresponding cluster masses used for 
Figs.~\ref{fig:mxml} and \ref{fig:myml}.
All masses are in units of $10^{14}\,\mathrm{M}_{\odot}$, without applying the $E(z)$
factor. We state only stochastic uncertainties, i.e.\ do not include systematics.
The proxy-based masses $M_{500}^{\mathrm{Y,T,G}}$ from V09a in the first part of this Table are 
defined in Sect.~\ref{sec:massdef}, as well as the $r_{500}^{\mathrm{Y,T,G}}$ quoted in the second 
part. The third and fourth part of the Table show the weak lensing ($M^{\mathrm{wl}}$) and hydrostatic
($M^{\mathrm{hyd}}$) for different cases, respectively. By $c_{\mathrm{fit}}$ and $c_{\mathrm{B13}}$
we denote the choices for the NFW concentration parameter explained in Sect.~\ref{sec:wla}. We refer
to Sect.~\ref{sec:ndc} for the introduction of the ``no dilu.\ corr.'' case (only differing from the
default for the first four clusters). See Table~\ref{tab:amasses} for further properties.}
  \renewcommand{\arraystretch}{1.1}
 \renewcommand\tabcolsep{3.5pt}
 \begin{center}
  \begin{tabular}{c|cccccccc} \hline\hline
 & CL\,0030 & CL\,0159 & CL\,0230 & CL\,0809 & CL\,1357 & CL\,1416 & CL\,1641 & CL\,1701 \\
 & +2618 & +0030 & +1836 & +2811 & +6232 & +4446 & +4001 & +6414 \\ \hline
   Redshift $z$ & $0.50$ & $0.39$ & $0.80$ & $0.40$ & $0.53$ & $0.40$ & $0.46$ & $0.45$ \\
 $k_{\mathrm{B}} T_{\mathrm{X}}$ from V09a [$\mbox{keV}$]& $5.63\pm1.13$ & $4.25\pm0.96$ & $5.50\pm1.02$ & $4.17\pm0.73$ & $4.60\pm0.69$ & $3.26\pm0.46$ & $3.31\pm0.62$ & $4.36\pm0.46$ \\
 $M_{500}^{\mathrm{Y}}$ from V09a & $3.43\pm0.41$ & $2.51\pm0.37$ & $3.46\pm0.46$ & $3.69\pm0.42$ & $2.96\pm0.29$ & $2.52\pm0.24$ & $1.70\pm0.20$ & $3.28\pm0.24$ \\ 
 $M_{500}^{\mathrm{T}}$ from V09a & $4.41\pm1.33$ & $2.67\pm0.90$ & $3.57\pm0.99$ & $2.96\pm0.78$ & $2.78\pm0.62$ & $1.76\pm0.37$ & $1.73\pm0.49$ & $2.66\pm0.42$ \\ 
 $M_{500}^{\mathrm{G}}$ from V09a & $2.04\pm0.19$ & $1.92\pm0.22$ & $2.70\pm0.27$ & $3.98\pm0.35$ & $2.40\pm0.18$ & $3.10\pm0.24$ & $1.34\pm0.13$ & $3.20\pm0.20$ \\ \hline
 $r_{200}^{\mathrm{wl}}(c_{\mathrm{fit}})$ [kpc] from Paper II & $1520_{-160}^{+140}$ & $1370_{-220}^{+180}$ & $1540_{-320}^{+280}$ & $1750_{-280}^{+230}$ & $1110_{-250}^{+210}$ & $980_{-180}^{+150}$ & $1060_{-260}^{+300}$ & $940_{-290}^{+320}$ \\
 $r_{200}^{\mathrm{wl}}(c_{\mathrm{B13}})$ [kpc] from Paper II& $1445_{-151}^{+132}$ & $1312_{-275}^{+222}$ & $1514_{-317}^{+273}$ & $1714_{-221}^{+192}$ & $1057_{-236}^{+202}$ & $977_{-183}^{+155}$ & $1014_{-126}^{+191}$ & $1014_{-201}^{+166}$ \\
 $r_{200}^{\mathrm{wl}}(c_{\mathrm{fit}})$, no dilu.\ corr.\ [kpc] & $1430_{-150}^{+140}$ & $1320_{-200}^{+170}$ & $1470_{-310}^{+280}$ & $1660_{-270}^{+220}$ & -- & -- & -- & -- \\
 $r_{200}^{\mathrm{wl}}(c_{\mathrm{B13}})$, no dilu.\ corr.\ [kpc] & $1349_{-136}^{+123}$ & $1265_{-256}^{+214}$ & $1432_{-305}^{+258}$ & $1629_{-204}^{+182}$ & -- & -- & -- & -- \\ \hline
 $r_{500}^{\mathrm{wl}}(c_{\mathrm{fit}})$ [kpc] & $914_{-96}^{+84}$ & $959_{-154}^{+126}$ & $974_{-203}^{+177}$ & $994_{-159}^{+131}$ & $674_{-152}^{+127}$ & $651_{-120}^{+100}$ & $440_{-108}^{+125}$ & $404_{-125}^{+137}$ \\
 $r_{500}^{\mathrm{wl}}(c_{\mathrm{B13}})$ [kpc] & $933_{-96}^{+84}$ & $848_{-175}^{+142}$ & $959_{-203}^{+171}$ & $1108_{-142}^{+129}$ & $685_{-155}^{+129}$ & $637_{-117}^{+104}$ & $655_{-182}^{+136}$ & $655_{-130}^{+110}$ \\
 $r_{500}^{\mathrm{wl}}(c_{\mathrm{fit}})$, no dilu.\ corr.\ [kpc] & $850_{-89}^{+83}$ & $924_{-140}^{+119}$ & $919_{-194}^{+175}$ & $923_{-150}^{+122}$ & -- & -- & -- & -- \\
 $r_{500}^{\mathrm{wl}}(c_{\mathrm{B13}})$, no dilu.\ corr.\ [kpc] & $870_{-90}^{+87}$ & $816_{-168}^{+136}$ & $909_{-191}^{+165}$ & $1051_{-129}^{+116}$ & -- & -- & -- & -- \\
 $r_{500}^{\mathrm{Y}}$ [kpc] & $873\pm85$ & $821\pm109$ & $777\pm75$ & $930\pm84$ & $821\pm92$ & $819\pm108$ & $702\pm138$ & $877\pm89$ \\
 $r_{500}^{\mathrm{T}}$ [kpc] & $949\pm72$ & $838\pm105$ & $785\pm73$ & $864\pm97$ & $804\pm96$ & $727\pm138$ & $706\pm136$ & $818\pm102$ \\
 $r_{500}^{\mathrm{G}}$ [kpc] & $734\pm120$ & $751\pm130$ & $715\pm88$ & $954\pm80$ & $766\pm106$ & $877\pm94$ & $648\pm161$ & $870\pm91$ \\ \hline
 $M_{500}^{\mathrm{wl}}(r_{500}^{\mathrm{wl}})$, using $c_{\mathrm{fit}}$ & $3.94_{-1.12}^{+1.19}$ & $4.00_{-1.63}^{+1.79}$ & $6.82_{-3.44}^{+4.43}$ & $4.51_{-1.84}^{+2.03}$ & $1.64_{-0.88}^{+1.11}$ & $1.27_{-0.58}^{+0.68}$ & $0.42_{-0.24}^{+0.47}$ & $0.32_{-0.22}^{+0.45}$ \\
 $M_{500}^{\mathrm{wl}}(r_{500}^{\mathrm{wl}})$, using $c_{\mathrm{B13}}$ & $4.19_{-1.16}^{+1.23}$ & $2.77_{-1.38}^{+1.64}$ & $6.51_{-3.32}^{+4.14}$ & $6.24_{-2.11}^{+2.44}$ & $1.72_{-0.92}^{+1.16}$ & $1.19_{-0.54}^{+0.68}$ & $1.38_{-0.86}^{+1.05}$ & $1.37_{-0.66}^{+0.81}$ \\
 $M_{500}^{\mathrm{wl}}(r_{500}^{\mathrm{wl}})$, using $c_{\mathrm{fit}}$, no dilu.\ corr.\ & $3.17_{-0.89}^{+1.02}$ & $3.58_{-1.39}^{+1.57}$ & $5.73_{-2.92}^{+3.93}$ & $3.61_{-1.49}^{+1.63}$ & -- & -- & -- & -- \\
 $M_{500}^{\mathrm{wl}}(r_{500}^{\mathrm{wl}})$, using $c_{\mathrm{B13}}$, no dilu.\ corr.\ & $3.40_{-0.95}^{+0.98}$ & $2.46_{-1.23}^{+1.45}$ & $5.54_{-2.81}^{+3.60}$ & $5.33_{-1.73}^{+1.97}$ & -- & -- & -- & -- \\
 $M_{500}^{\mathrm{wl}}(r_{500}^{\mathrm{Y}})$, using $c_{\mathrm{B13}}$ & $3.91_{-0.83}^{+0.78}$ & $2.68_{-1.12}^{+0.93}$ & $5.15_{-2.07}^{+1.83}$ & $5.15_{-1.34}^{+1.25}$ & $2.04_{-0.96}^{+0.81}$ & $1.49_{-0.59}^{+0.52}$ & $1.48_{-0.85}^{+0.64}$ & $1.78_{-0.76}^{+0.64}$ \\
 $M_{500}^{\mathrm{wl}}(r_{500}^{\mathrm{T}})$, using $c_{\mathrm{B13}}$ & $4.26_{-0.91}^{+0.83}$ & $2.73_{-1.15}^{+0.95}$ & $5.21_{-2.10}^{+1.85}$ & $4.76_{-1.24}^{+1.17}$ & $2.00_{-0.94}^{+0.79}$ & $1.34_{-0.52}^{+0.47}$ & $1.48_{-0.85}^{+0.64}$ & $1.68_{-0.71}^{+0.60}$ \\
 $M_{500}^{\mathrm{wl}}(r_{500}^{\mathrm{G}})$, using $c_{\mathrm{B13}}$ & $3.26_{-0.70}^{+0.69}$ & $2.45_{-1.02}^{+0.85}$ & $4.68_{-1.86}^{+1.66}$ & $5.29_{-1.38}^{+1.28}$ & $1.91_{-0.89}^{+0.75}$ & $1.58_{-0.63}^{+0.56}$ & $1.37_{-0.78}^{+0.59}$ & $1.77_{-0.75}^{+0.64}$ \\ \hline
 $M_{500}^{\mathrm{hyd}}(r_{500}^{\mathrm{wl}})$, using $c_{\mathrm{fit}}$ & $3.46_{-1.01}^{+1.10}$ & $2.78_{-1.00}^{+1.11}$ & $5.13_{-2.37}^{+2.95}$ & $3.66_{-1.26}^{+1.36}$ & $2.20_{-0.85}^{+0.92}$ & $1.31_{-0.44}^{+0.51}$ & $1.17_{-0.51}^{+0.70}$ & $1.02_{-0.45}^{+0.60}$ \\
 $M_{500}^{\mathrm{hyd}}(r_{500}^{\mathrm{wl}})$, using $c_{\mathrm{B13}}$ & $3.47_{-1.00}^{+1.10}$ & $2.50_{-1.01}^{+1.12}$ & $4.99_{-2.33}^{+2.84}$ & $3.99_{-1.22}^{+1.35}$ & $2.20_{-0.85}^{+0.91}$ & $1.28_{-0.43}^{+0.51}$ & $1.65_{-0.72}^{+0.75}$ & $1.70_{-0.52}^{+0.53}$ \\
 $M_{500}^{\mathrm{hyd}}(r_{500}^{\mathrm{wl}})$, using $c_{\mathrm{fit}}$, no dilu. corr.\ & $3.21_{-0.94}^{+1.05}$ & $2.68_{-0.94}^{+1.06}$ & $4.66_{-2.18}^{+2.81}$ & $3.35_{-1.18}^{+1.27}$ & -- & -- & -- & -- \\
 $M_{500}^{\mathrm{hyd}}(r_{500}^{\mathrm{wl}})$, using $c_{\mathrm{B13}}$, no dilu. corr.\ & $3.22_{-0.94}^{+1.02}$ & $2.40_{-0.97}^{+1.08}$ & $4.54_{-2.12}^{+2.64}$ & $3.75_{-1.13}^{+1.26}$ & -- & -- & -- & -- \\
 \hline\hline
 \end{tabular} 
  \label{tab:masses}
 \end{center}
\end{table*}
We present only a brief description of the WL analysis in this paper; for more
details see Paper~II. Basic data reduction is performed
using the \texttt{THELI} pipeline for multi-chip cameras \citep{2005AN....326..432E}, 
adapted to MMT/Megacam. We employ the photometric calibration by \citet{2006A&A...452.1121H}.  
Following \citet{2007A&A...470..821D}, regions of the \texttt{THELI} coadded images not suitable
for WL shear measurements are masked.
Shear is measured using an implementation of the ``KSB+'' algorithm 
\citep{1995ApJ...449..460K,2001A&A...366..717E}, the ``TS'' pipeline
\citep{2006MNRAS.368.1323H,2007A&A...468..823S,2009A&A...504..689H}.
Catalogues of lensed background galaxies are selected based on the available colour information.
For clusters covered in three filters, we include galaxies based on their position in 
colour-colour-magnitude space (Paper~II; see Klein et al., in prep., for a generalisation).
For clusters covered only in one passband, we apply a magnitude cut.
Where available, colour information also enables us to quantify and correct for the
dilution by residual cluster members \citep{2007MNRAS.379..317H} in the shear catalogues.
The mass normalisation of the WL signal is set by the mean lensing depth 
$\langle\beta\rangle$, defined as $\beta\!=\!D_{\mathrm{ds}}/D_{\mathrm{s}}$, the ratio of angular 
diameter distances between the deflector and the source, and between the observer and the source. 
The \citet{2006A&A...457..841I} CFHTLS Deep fields photometric redshift catalogue serves as a
proxy for estimating $\langle\beta\rangle$ and for calibrating the background selection. 

The tangential ellipticity profiles given the \textit{Rosat} cluster centres are modelled by 
fitting the reduced shear profile \citep{1996A&A...313..697B,2000ApJ...534...34W} corresponding 
to the \citet[NFW]{1996ApJ...462..563N,1997ApJ...490..493N} density profile between 
$0.2\,\mbox{Mpc}$ and $5.0\,\mbox{Mpc}$ projected radius. Input ellipticities are scaled
according to the \citet{2009A&A...504..689H} calibration factor and, where applicable, with the
correction for dilution by cluster members. We consider the intrinsic source ellipticity 
measured from the data, accounting for its dependence on the shear \citep{2000A&A...353...41S}.

Lensing masses are inferred by evaluating a $\chi^{2}$  merit function 
on a grid in radius $r_{200}$ and concentration $c_{200}$. 
The latter is poorly constrained in the direct fit, so  we marginalise
over it assuming an empirical mass-concentration relation. In addition to  the direct fit approach,
in \citet{2012A&A...546A..79I}, we report masses using two different mass-concentration 
relations: \citet[B01]{2001MNRAS.321..559B}, and \citet[B13]{2013ApJ...766...32B}\footnote{
Actually, we use the slightly different relation as given in Version~1, referred to in Paper II
as ``B12'': \texttt{arxiv.org/abs/1112.5479v1}.}. Finding the masses using B01 or B13 less
susceptible to variations in the model in Paper~II, we explore their effect further in 
Sect.~\ref{sec:c200}.

\subsection{Choice of the overdensity contrast}

Cluster scaling relations are usually given for the mass contained within a 
radius $r_{500}$, corresponding to an overdensity  
$\Delta\!=\!500$ compared to the \emph{critical} density $\rho_{\mathrm{c}}$
of the Universe at the cluster redshift. 
This $\Delta$ is chosen  because the best-constrained X-ray masses are found 
close to $r_{500}$, determined by the particle backgrounds  of \textit{Chandra} and \textit{XMM-Newton}
\citep[cf.][]{2010PASJ...62..811O}. Currently, only \textit{Suzaku} allows direct 
constraints upon X-ray 
masses at $r_{200}$ \citep[see][and references therein]{2013SSRv..177..195R}.
In order to compare to the results from the \citet{2009ApJ...692.1033V} \textit{Chandra} 
analysis, we compute our $\Delta\!=\!500$ WL masses from our $\Delta\!=\!200$ masses, 
assuming the fitted NFW profiles  given by $(r_{200},c_{200})$  to be correct.
Independent of $\Delta\!>\!1$, the cumulative mass of a NFW halo, described by $r_{\Delta}$ and 
$c_{\Delta}$, out to a test radius $r$ is given by:
\begin{align} \label{eq:mnfw}
M_{\mathrm{NFW}}(r)&=&\Delta\rho_{\mathrm{c}}\frac{4\pi}{3}r_{\Delta}^{3} &\times&
\frac{\ln{(1\!+\!c_{\Delta}r/r_{\Delta})}-\frac{c_{\Delta}r/r_{\Delta}}{1\!+\!c_{\Delta}r/r_{\Delta}}}
{\ln{(1\!+\!c_{\Delta})}-c_{\Delta}/(1\!+\!c_{\Delta})}\\
&=& M_{\Delta}(r_{\Delta}) &\times& \Xi(r;r_{\Delta},c_{\Delta}),
\end{align} 
separating into the mass $M_{\Delta}$ and a function we call $\Xi(r;r_{\Delta},c_{\Delta})$. 
Equating Eq.~(\ref{eq:mnfw}) with $r\!=\!r_{500}$ for $\Delta\!=\!200$ and $\Delta'\!=\!500$,
we arrive at this implicit equation for $r_{500}$, which we solve numerically:
\begin{equation} \label{eq:xi}
r_{500}=r_{200}\left(\tfrac{2}{5}\,\Xi(r_{500},r_{200},c_{200})\right)^{1/3}.
\end{equation}

\subsection{X-ray analysis} \label{sec:xray}
Under the strong assumptions that the ICM is in hydrostatic equilibrium
and follows a spherically symmetric mass distribution,
the cluster mass within a radius $r$ can be calculated as 
\citep[see e.g.][]{1988xrec.book.....S}:
\begin{equation} \label{eq:xmass}
M^{\mathrm{hyd}}(r)\!=\!\frac{-k_{\mathrm{B}}T_{\mathrm{X}}(r)\,r}{\mu m_{\mathrm{p}}G}
\left(\frac{\mathrm{d}\ln \rho_{\mathrm{g}}(r)}{\mathrm{d}\ln r} + 
\frac{\mathrm{d}\ln T_{\mathrm{X}}(r)}{\mathrm{d}\ln r}\right)
\end{equation}
from the ICM density and temperature profiles $\rho_{\mathrm{g}}(r)$ 
and $T_{\mathrm{X}}(r)$,  where $G$ is the gravitational constant, 
$m_{\mathrm{p}}$ is the proton mass, and $\mu=0.5954$ the mean molecular mass of the ICM.
The ICM density is modelled by fitting the observed \textit{Chandra}
surface brightness profile, assuming  a primordial He
abundance and an ICM metallicity of $0.2\,\mathrm{Z}_{\sun}$, 
such that $\rho_{\mathrm{g}}(r)\!=\!1.274\,m_{\mathrm{p}}\,n(r)$.
We use a \citet{2006ApJ...640..691V} particle density profile with 
$n(r)\!=\!\!\sqrt{n_{\mathrm{p}}(r)\,n_{\mathrm{e}}(r)}$. Extending  the 
widely-used $\beta$-profile \citep{1978A&A....70..677C}, it allows for prominent
cluster cores as well as steeper surface brightness profiles in the cluster outskirts to be 
modelled by additional terms. 

The relatively low signal/noise in the \textit{Chandra} data renders the
determination of individual temperature profiles difficult. Rather, we fit a \emph{global}
$T_{\mathrm{X}}$ (Table~\ref{tab:masses}; V09a) and assume the empirical average temperature profile
$T_{\mathrm{X}}(r)\!=\!T_{\mathrm{X}}\left(1.19-0.84r/r_{200}\right)$
\citet{2013SSRv..177..195R} derive from compiling all available \textit{Suzaku} 
temperature profiles (barring only the two most exceptional clusters). 
For $r_{200}$, we use the WL results from Paper~II.\footnote{For the 
scaling relations within WL-derived radii, we choose the respective 
$c_{\mathrm{NFW}}$. Otherwise, we use $c_{\mathrm{B13}}$ as a default.} 

Equation~(\ref{eq:xmass}) provides us with a cumulative mass profile. 
We evaluate this profile at some $r_{\mathrm{test}}$, e.g.\ from WL, and propagate
the uncertainty in $r_{\mathrm{test}}$, together with the uncertainty in $T_{\mathrm{X}}$.

Hydrostatic equilibrium and sphericity are known to be problematic assumptions
for many clusters. Nonetheless, hydrostatic masses are commonly used in the literature
in comparisons to WL masses. Our goal is to study if and how biases due to deviations 
from the above-mentioned assumptions show up.

\subsection{Mass Estimates} \label{sec:massdef}

Table~\ref{tab:masses} comprises the key results on radii $r_{500}$ and the corresponding
mass estimates. By $M^{\mathcal{P}}(r^{\mathcal{Q}})$, we denote a mass measured from data on
proxy $\mathcal{P}$ within a radius defined by proxy $\mathcal{Q}$. We use five mass estimates:
$\mathcal{P,Q}\in\{\mathrm{wl,hyd,Y,T,G}\}$. The first two are the weak lensing (wl) and hydrostatic
X-ray masses (hyd), as introduced in Sects.~\ref{sec:wla} and \ref{sec:xray}. 
Having analysed deep \textit{Chandra} observations they acquired,
\citet{2009ApJ...692.1033V} present three further mass estimates for all $36$ clusters 
in the complete sample. Based on the proxies 
$T_{\mathrm{X}}$, the ICM mass $M_{\mathrm{gas}}$, and 
$Y_{\mathrm{X}}\!=\!T_{\mathrm{X}}M_{\mathrm{gas}}$, mass estimates
$M^{\mathrm{T}}$, $M^{\mathrm{G}}$, and $M^{\mathrm{Y}}$
are quoted in Table~2 of V09a. We point out that V09a obtain these estimates
by calibrating the mass scaling relations for respective proxy on local clusters 
(see their Table~3). V09a further provide a detailed account of 
the relevant systematic sources of uncertainty. 

The radii 
$r_{500}^{\mathcal{P}}\!=\!\left(3M^{\mathcal{P}}_{500}/(2000\pi\rho_{\mathrm{c}})\right)^{1/3}$ 
listed in Table~\ref{tab:masses} are obtained from 
$M_{500}^{\mathcal{P}},\,\mathcal{P}\!\in\!\{\mathrm{Y,T,G}\}$.
Using Eqs.~(\ref{eq:mnfw}) and (\ref{eq:xmass}), we then derive the
WL and hydrostatic masses, respectively, within these radii.
We emphasise that all
WL mass uncertainties quoted in Table~\ref{tab:masses} are purely statistical and do not include
any of the systematics discussed in Paper~II.

\subsection{Fitting algorithm for scaling relations} \label{sec:regress}

The problem of selecting the best linear representation $y=A+Bx$ for a sample of (astronomical)
observations of two quantities $\{x_{i}\}$ and $\{y_{i}\}$ can be surprisingly complex. 
A plethora of algorithms and literature cope  with the different assumptions
about measurement uncertainties one can or has to make 
\citep[e.g.][]{1992nrca.book.....P,1996ApJ...470..706A,2002ApJ...574..740T,2007ApJ...665.1489K,2010arXiv1008.4686H,2010MNRAS.409.1330W,2012arXiv1210.6232A,feigelson2012modern}.
The challenges observational astronomers have to tackle 
when trying to reconcile the prerequisites
of statistical estimators with the realities of astrophysical data are manifold, including
heteroscedastic uncertainties (i.e.\ depending non-trivially on the data themselves), intrinsic
scatter, poor knowledge of systematics, poor sample statistics, ``outlier'' points, and 
non-Gaussian probability distributions.
Tailored to the problem of galaxy cluster scaling relations, \citet{2012arXiv1212.0858M} 
proposed a ``self-consistent'' modelling approach based on the fundamental observables. 
A full account of these different effects exceeds the scope of this article. 
We  choose the relatively simple \texttt{fitexy} algorithm \citep{1992nrca.book.....P}, 
minimising the estimator
\begin{equation} \label{eq:fitexy}
\chi^{2}_{\mathrm{P92}} = \sum_{i=1}^{N}{\frac{\left(y_{i}-A-Bx_{i}\right)^{2}}
{\sigma_{y,i}^{2}+B^{2}\sigma_{x,i}^{2}}}\quad,
\end{equation}
which allows the uncertainties $\sigma_{x,i}$ and $\sigma_{y,i}$ to vary for different data 
points $x_{i}$ and $y_{i}$, but assumes them to be drawn from a Gaussian distribution.
To accommodate intrinsic scatter, $\sigma_{y,i}^{2}$ in Eq.~(\ref{eq:fitexy}) can be 
replaced by $\sigma_{i}^{2}\!=\!\sigma_{y,i}^{2}+\sigma_{\mathrm{int}}^{2}$
\citep[e.g.][]{2006ApJ...653.1049W,2012arXiv1210.6232A}.
We test for intrinsic scatter using \texttt{mpfitexy} \citep{2009ASPC..411..251M,2010MNRAS.409.1330W},
but in most cases, due to the small $\chi^{2}$ values, find the respective parameter not invoked.
Thus we decide against this additional complexity. 
A strength of Eq.~(\ref{eq:fitexy}) is its invariance
under changing $x$ and $y$ \citep[e.g.][]{2002ApJ...574..740T}; i.e., we do not assume either to
be ``the independent variable''.

Rather than propagating  the (unknown) distribution functions in the mass
uncertainties\footnote{A natural feature in complex measurements like this, 
asymmetric uncertainties in $M_{200}^{\mathrm{wl}}$ arise from the grid
approach to $\chi^{2}$ minimisation in Paper~II (cf.~Fig.~2 therein)}, we approximate 
$1\sigma$ Gaussian uncertainties \emph{in decadic log-space}, applying the  
symmetrisation:
\begin{equation}  \label{eq:sigmalog}
\sigma_{(\log{\xi_{i}})} = \log{(\mathrm{e})}\cdot(\xi_{i}^{+}\!-\!\xi_{i}^{-})/(2\xi_{i}) 
= \log{(\mathrm{e})}\cdot(\sigma^{+}_{\!\xi,i}\!+\!\sigma^{-}_{\!\xi,i})/(2\xi_{i})\,, 
\end{equation}   
where $\xi_{i}^{+}\!=\!\xi_{i}\!+\!\sigma^{+}_{\!\xi,i}$ and 
$\xi_{i}^{-}\!=\!\xi_{i}\!-\!\sigma^{-}_{\!\xi,i}$ are the upper and lower limits 
of the $1\sigma$ interval (in linear space) for the datum $\xi_{i}$, given 
the uncertainties $\sigma^{\pm}_{\!\xi,i}$. All our calculations 
and plots use $\{x_{i}\}\!:=\!\{\log{\xi_{i}}\}$ and $\{\sigma_{x,i}\}\!:=\!\{\sigma_{(\log{\xi_{i}})}\}$,
with $\log\!\equiv\!\log_{10}$.

\section{Results} \label{sec:res}

\subsection{Weak lensing and hydrostatic masses} \label{sec:xl}

\begin{table*}
 \caption{Measurements of the X-ray -- WL mass bias. 
We estimate a possible bias between masses $\xi$ and $\eta$ by three estimators: 
First, we fit to $(\log \xi -\log \eta)$ as a function of $\eta$, yielding an intercept $A$ at pivot
$\log{\left(M_{\mathrm{piv}}/\mathrm{M}_{\odot}\right)}\!=\!14.5$ and slope $B$ 
from the Monte Carlo/jackknife analysis. Second, we compute the logarithmic bias 
$b_{\mathrm{MC}}\!=\!\langle\log{\xi}\!-\!\log{\eta}\rangle_{\mathrm{MC}}$, averaged over the same realisations.
Uncertainties for the MC results are given by $1\sigma$ ensemble dispersions.
In parentheses next to $b_{\mathrm{MC}}$, we show its value for the low-$M^{\mathrm{wl}}$ 
and high-$M^{\mathrm{wl}}$ clusters.
Third, we quote the logarithmic bias $b\!=\!\langle\log{\xi}\!-\!\log{\eta}\rangle$ obtained directly
from the input masses, along with its standard error.
Finally, we give the $\chi^{2}_{\mathrm{red}}$ for the mass-mass scaling, obtained from the MC method. 
The ``default'' model denotes WL and hydrostatic masses as described in Sect.~\ref{sec:obsdat}.}
 \renewcommand{\arraystretch}{1.1}
 \renewcommand\tabcolsep{3pt}
  \begin{tabular}{ccc|ccccc|c}\hline\hline
    Scaling Relation & Model & $c_{\mathrm{NFW}}$ & Slope $B$ & Intercept $A$ & $b_{\mathrm{MC}}$ from Monte Carlo & $b\!=\!\langle\log{\xi}\!-\!\log{\eta}\rangle$ &  $\chi^{2}_{\mathrm{red,M-M}}$ &Section \\ \hline
    $M^{\mathrm{wl}}_{500}(r_{500}^{\mathrm{wl}})$--$M^{\mathrm{hyd}}_{500}(r_{500}^{\mathrm{wl}})$ & 
    default & $c_{\mathrm{fit}}$ & $-0.51_{-0.21}^{+0.20}$ & $0.00_{-0.08}^{+0.07}$  & $0.08_{-0.13}^{+0.14}$ ($0.27_{-0.20}^{+0.21}$; $-0.10_{-0.15}^{+0.16}$) & $0.08\pm0.09$ & $0.58$ & \ref{sec:xl} \\  
    & default & $c_{\mathrm{B13}}$ &  $-0.47_{-0.25}^{+0.26}$&$-0.02_{-0.08}^{+0.07}$  & $0.00_{-0.13}^{+0.14}$ ($0.10_{-0.18}^{+0.23}$; $-0.10_{-0.15}^{+0.17}$) & $-0.02\pm0.04$ & $0.52$ & \ref{sec:xl} \\ 
    & no dilu.\ corr.\ & $c_{\mathrm{fit}}$ &$-0.53\pm0.23$  & $0.01\pm0.08$ & $0.11_{-0.13}^{+0.14}$ ($0.27_{-0.20}^{+0.21}$; $-0.06_{-0.15}^{+0.16}$)  & $0.11\pm0.08$ & $0.57$ & \ref{sec:ndc}\\  
    & no dilu.\ corr.\ & $c_{\mathrm{B13}}$ &$-0.49\pm0.29$  & $-0.01_{-0.08}^{+0.07}$  & $0.02_{-0.13}^{+0.14}$ ($0.10_{-0.18}^{+0.22}$; $-0.06_{-0.15}^{+0.17}$)   & $0.00\pm0.03$ & $0.51$ & \ref{sec:ndc}\\ \hline
    $M^{\mathrm{wl}}(r_{\mathrm{fix}})$--$M^{\mathrm{hyd}}(r_{\mathrm{fix}})$ & $r_{\mathrm{fix}}\!=\!600\,\mbox{kpc}$ & $c_{\mathrm{B13}}$ & $-0.68_{-0.21}^{+0.19}$  & $-0.11\pm0.05$ & $0.01_{-0.07}^{+0.10}$ ($0.12_{-0.10}^{+0.16}$; $-0.11_{-0.08}^{+0.10}$)  & $-0.02\pm0.04$ & $0.82$ & \ref{sec:chisqdisc} \\
    & $r_{\mathrm{fix}}\!=\!800\,\mbox{kpc}$ & $c_{\mathrm{B13}}$ & $-0.58_{-0.21}^{+0.19}$ & $-0.02\pm0.04$ & $0.02_{-0.07}^{+0.10}$ ($0.12_{-0.11}^{+0.18}$; $-0.09_{-0.08}^{+0.10}$) & $-0.02\pm0.04$ & $0.72$ & \ref{sec:chisqdisc} \\
    & $r_{\mathrm{fix}}\!=\!1000\,\mbox{kpc}$ & $c_{\mathrm{B13}}$ & $-0.52_{-0.21}^{+0.19}$ & $0.01\pm0.05$ & $0.01_{-0.08}^{+0.11}$ ($0.10_{-0.11}^{+0.20}$; $-0.09_{-0.09}^{+0.11}$)  & $-0.03\pm0.03$ & $0.69$ & \ref{sec:chisqdisc} \\ \hline
    $M^{\mathrm{wl}}_{500}(r_{500}^{\mathrm{Y}})$--$M^{\mathrm{Y}}_{500}(r_{500}^{\mathrm{Y}})$ & 
     default & $c_{\mathrm{B13}}$ & $-0.75_{-0.13}^{+0.12}$ & $0.07\pm0.03$ & $0.08_{-0.07}^{+0.10}$ ($0.23_{-0.11}^{+0.18}$; $-0.08_{-0.07}^{+0.10}$)  & $0.04\pm0.06$ & $1.21$ & \ref{sec:ygt} \\
 $M^{\mathrm{wl}}_{500}(r_{500}^{\mathrm{T}})$--$M^{\mathrm{T}}_{500}(r_{500}^{\mathrm{T}})$ &
     default & $c_{\mathrm{B13}}$ & $-0.63\pm0.23$ & $0.04\pm0.06$  & $0.05_{-0.08}^{+0.11}$ ($0.17_{-0.12}^{+0.18}$; $-0.08_{-0.10}^{+0.11}$) & $0.02\pm0.05$ & $0.88$ & \ref{sec:ygt} \\
 $M^{\mathrm{wl}}_{500}(r_{500}^{\mathrm{G}})$--$M^{\mathrm{G}}_{500}(r_{500}^{\mathrm{G}})$ &
      default & $c_{\mathrm{B13}}$ & $-0.89_{-0.31}^{+0.18}$  & $0.01_{-0.04}^{+0.03}$ & $0.04_{-0.07}^{+0.10}$ ($0.21_{-0.10}^{+0.17}$; $-0.15_{-0.07}^{+0.09}$)  & $0.00\pm0.07$ & $2.11$ & \ref{sec:ygt}\\
   \hline\hline \label{tab:slopes}
  \end{tabular}
\end{table*}
\begin{figure}
 \includegraphics[width=8cm]{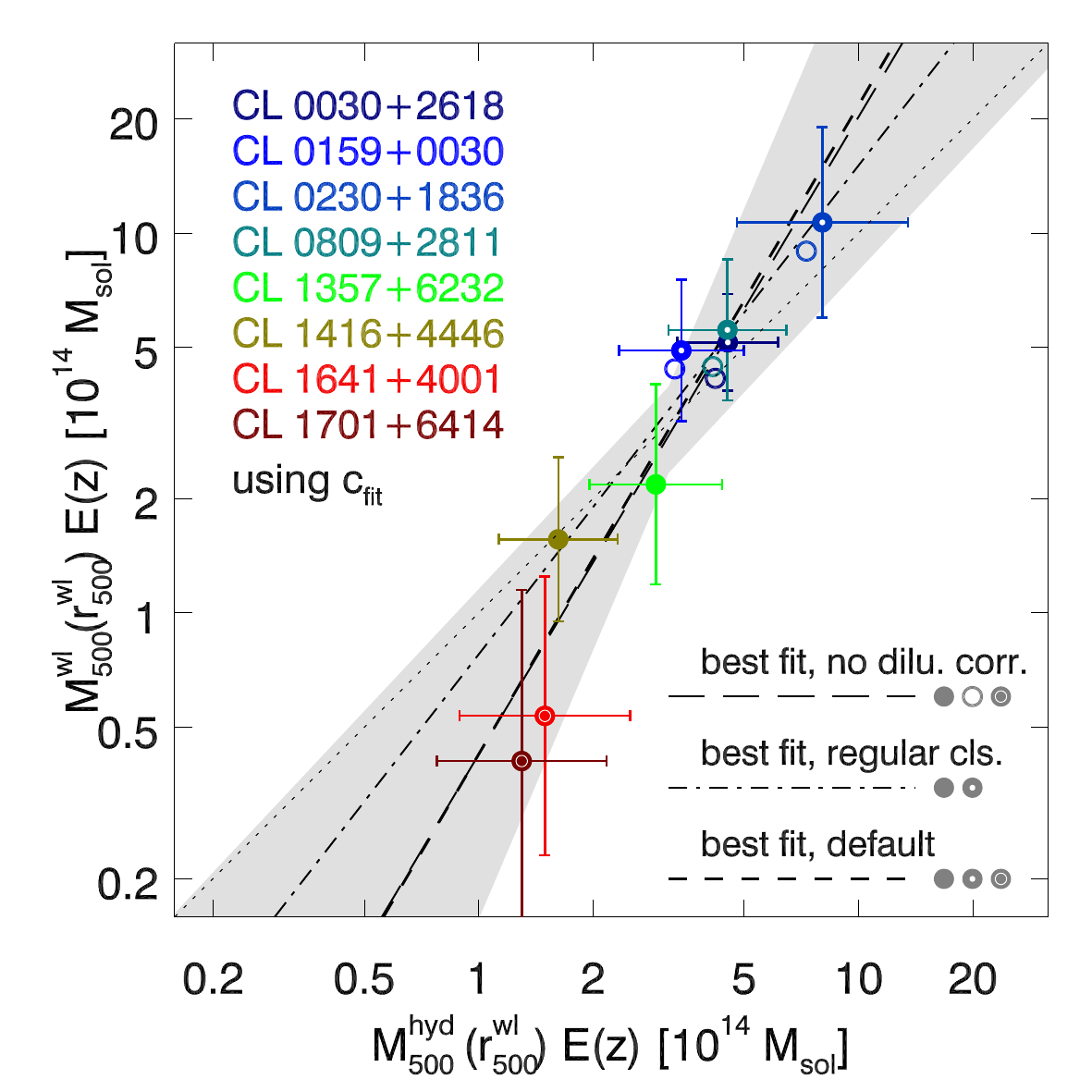}
 \includegraphics[width=8cm]{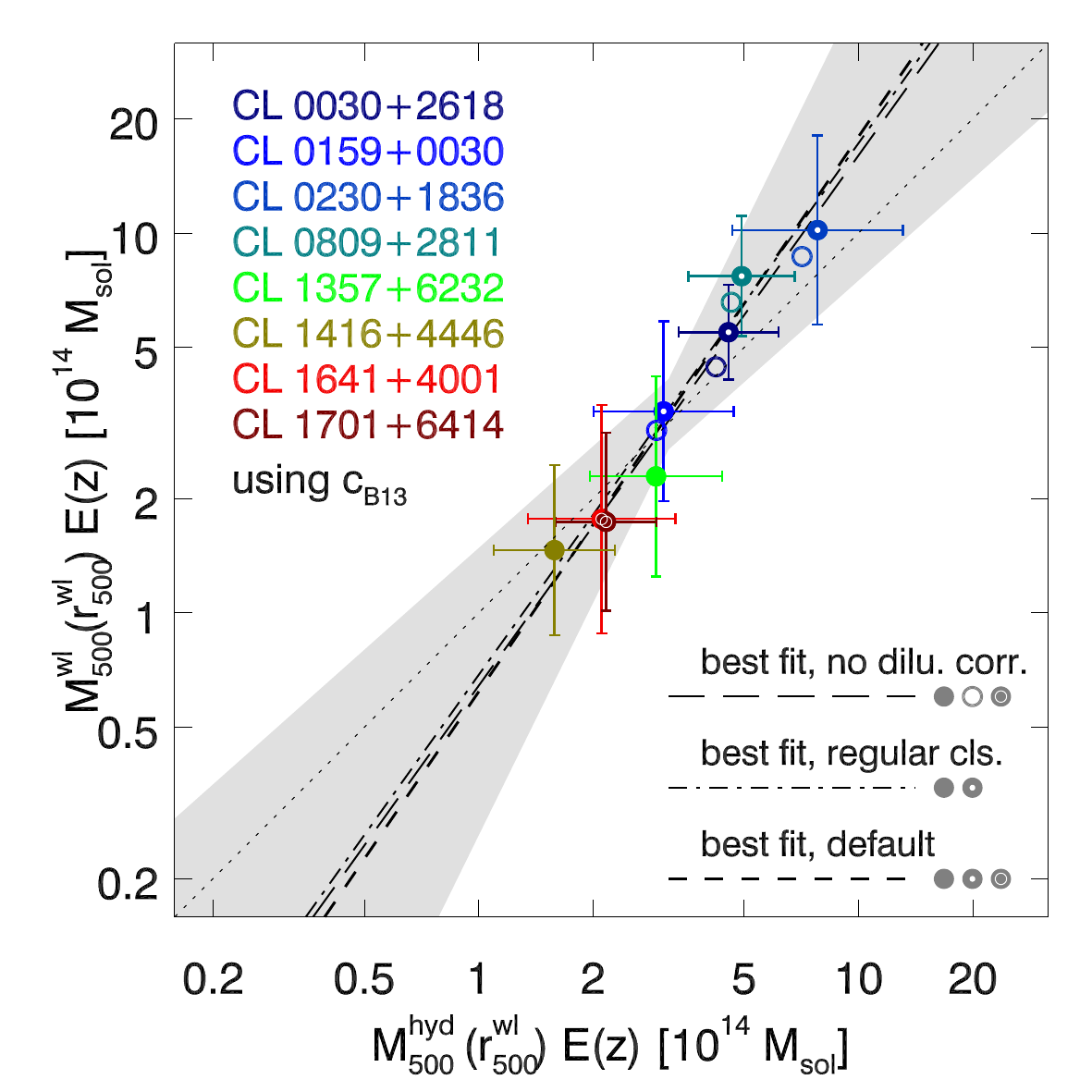}
 \caption{Scaling of weak lensing masses $M^{\mathrm{wl}}_{500}(r_{500}^{\mathrm{wl}})$ with 
  hydrostatic masses $M^{\mathrm{hyd}}_{500}(r_{500}^{\mathrm{wl}})$. The upper (lower) panel is for 
  $c_{\mathrm{fit}}$ ($c_{\mathrm{B13}}$). 
  Both show best fits for three cases: the default (filled, thick
  ring, dotted ring symbols; thick dashed line), regular shear profile clusters only (filled and
  thick ring symbols; dash-dotted line; Sect.~\ref{sec:c200}), 
  and without correction for dilution by cluster members (filled, thin ring, dotted ring symbols; 
  long dashed line; Sect.~\ref{sec:ndc}). The dotted line shows equality of the two 
  masses, $M^{\mathrm{wl}}_{500}\!\!=\!\!M^{\mathrm{hyd}}_{500}$. Shaded regions indicate the 
  uncertainty range of the default best-fit. Some error bars were omitted for sake of clarity.}
 \label{fig:mxml}
\end{figure}

The first and single most important observation is that hydrostatic masses
$M^{\mathrm{hyd}}_{500}(r_{500}^{\mathrm{wl}})$, i.e.\ evaluated at $r_{500}$ as found from weak
lensing, and weak lensing masses $M^{\mathrm{wl}}_{500}(r_{500}^{\mathrm{wl}})$ 
roughly agree with each other (Table~\ref{tab:masses}).
Our second key observation is the very tight scaling behaviour between $M^{\mathrm{hyd}}_{500}$ 
and $M^{\mathrm{wl}}_{500}$, as Fig.~\ref{fig:mxml} shows. In all cases presented in 
Fig.~\ref{fig:mxml}, and most of the ones we tested, all data points are consistent with the
best-fit relation. Consequently, the fits return small values of 
$\chi^{2}_{\mathrm{red}}\!<\!1$ (see Table~\ref{tab:slopes}). 
Bearing in mind that we only use stochastic uncertainties, this 
points to some intrinsic correlation of the WL and hydrostatic masses. 
We will discuss this point in Sect.~\ref{sec:chisqdisc}.

Finally, we find the slope of the 
$M^{\mathrm{wl}}_{500}(r_{500}^{\mathrm{wl}})$--$M^{\mathrm{hyd}}_{500}(r_{500}^{\mathrm{wl}})$ 
relation (dashed lines in Fig.~\ref{fig:mxml}) to be steeper than unity (dotted line):
Using the ``default model'', i.e.\ the analysis  described in Sect.~\ref{sec:obsdat}, a \texttt{fitexy} fit 
yields $1.71\pm0.64$ for the ``$c_{\mathrm{fit}}$'' case (concentration parameters from the shear profile fits, 
cf.\ Paper~II; upper panel of Fig.~\ref{fig:mxml}),
and $1.46\pm0.57$, if the B13 mass--concentration relation is applied
(``$c_{\mathrm{B13}}$''; lower panel).
The different slopes in the $c_{\mathrm{fit}}$ and $c_{\mathrm{B13}}$ cases are mainly due to 
the two clusters, CL\,1641+4001
and CL\,1701+6414, in which the weak lensing analysis revealed shallow tangential shear 
profiles due to extended surface mass plateaus (cf.\ Figs.~3 and 5 of Paper II). 
This will be the starting point for further analysis and interpretation in Sect.~\ref{sec:c200}. 

Although the $c_{\mathrm{B13}}$ slope is consistent with the expected $1$:$1$ relation, such a
$M^{\mathrm{wl}}_{500}$--$M^{\mathrm{hyd}}_{500}$ relation would translate to extreme
biases between X-ray and WL masses if extrapolated to higher and lower masses. 
Especially for masses of a few $10^{15}\,\mbox{M}_{\sun}$, ample observations disagree with the extrapolated
$M^{\mathrm{wl}}\!>\!2M^{\mathrm{hyd}}$.
We do not claim our data to have such predicting power outside its mass range.
Rather, we focus on what can be learnt about the X-ray/WL mass bias in our
$0.4\!\sim\!z\!\sim\!0.5$ mass range, which we, for the first time, study in the mass range
down to $\sim\!1\times10^{14}\,\mbox{M}_{\sun}$. 

We are using three methods to test for the biases between X-ray and
WL masses. First, we compute the logarithmic bias 
$b\!\!=\!\!\langle\log{\xi}\!-\!\log{\eta}\rangle$, which we
define as the average logarithmic difference between 
two general quantities $\xi$ and $\eta$. 
Its interpretation is that $10^{b}\eta$ is the average value corresponding to $\xi$.
The uncertainty in $b$ is given by the standard error of $(\log{\xi}\!-\!\log{\eta})$.
Hence, our measurement of $b\!\!=\!\!-0.02\pm0.04$ for $c_{\mathrm{B13}}$ 
corresponds to a vanishing fractional bias of
$\langle M^{\mathrm{hyd}}\rangle\approx(0.97\pm0.09)\langle M^{\mathrm{wl}}\rangle$. 

Given the small sample size, large uncertainties, and the tight scaling relations in Fig,~\ref{fig:mxml}
pointing to some correlation between the WL and X-ray masses, we base our further tests on a
Monte Carlo (MC) analysis including the jackknife test.
For $10^{5}$ realisations, we chose $\hat{\xi}_{i,k}\!=\!\xi_{i}\!+\!\delta\xi_{i,k}$
with random $\delta\xi_{i,k}$ drawn from zero-mean distributions 
assembled from two Gaussian halves with variances $\sigma^{-}_{\!\xi,i}$ for the negative and
$\sigma^{+}_{\!\xi,i}$ for the positive half.\footnote{Unphysical cluster masses
$<\!\!10^{13}\,\mbox{M}_{\sun}/E(z)$ are set to $10^{13}\,\mbox{M}_{\sun}/E(z)$.}
This provides a simple way of accommodating asymmetric uncertainties (cf.\ Paper~II and Table~\ref{tab:masses}).
Then we take the logarithm and again symmetrise the errors. 
We repeat for $\hat{\eta}_{i,k}\!=\!\eta_{i}\!+\!\delta\eta_{i,k}$.
On top, for each realisation $\{\hat{\xi}_{i,k},\hat{\eta}_{i,k}\}$, we discard one cluster
after another, yielding a total of  $8\!\times\!10^{5}$ samples. 

Based on those MC/jackknife realisations, we compute our second bias estimator
$b_{\mathrm{MC}}\!\!=\!\!\langle\log{\hat{\xi}}\!-\!\log{\hat{\eta}}\rangle_{\mathrm{MC}}$.
In order to achieve the best possible robustness against large uncertainties and small cluster numbers, 
we quote the ensemble median and dispersion.
We find $b_{\mathrm{MC}}\!=\!0.00_{-0.13}^{+0.14}$ for $c_{\mathrm{B13}}$, in good agreement
with $b\!\!=\!\!-0.02\pm0.04$, i.e. a median WL/X-ray mass ratio of $1$.

Fitting  $\log{(M^{\mathrm{X}}/M^{\mathrm{wl}})}$ as a function of $M^{\mathrm{wl}}$
and averaging over the MC/jackknife samples, we obtain our third bias estimator, an intercept $A$ 
at the pivot mass of $\log\left(M_{\mathrm{piv}}/\mbox{M}_{\sun}\right)\!=\!14.5$.
We find $A\!\!=\!\!-0.02_{-0.08}^{+0.07}$ for $c_{\mathrm{B13}}$, again consistent with vanishing bias.

\subsection{Lensing masses and X-ray masses from proxies} \label{sec:ygt}
\begin{figure*}
 \begin{center}
 \includegraphics[width=8cm]{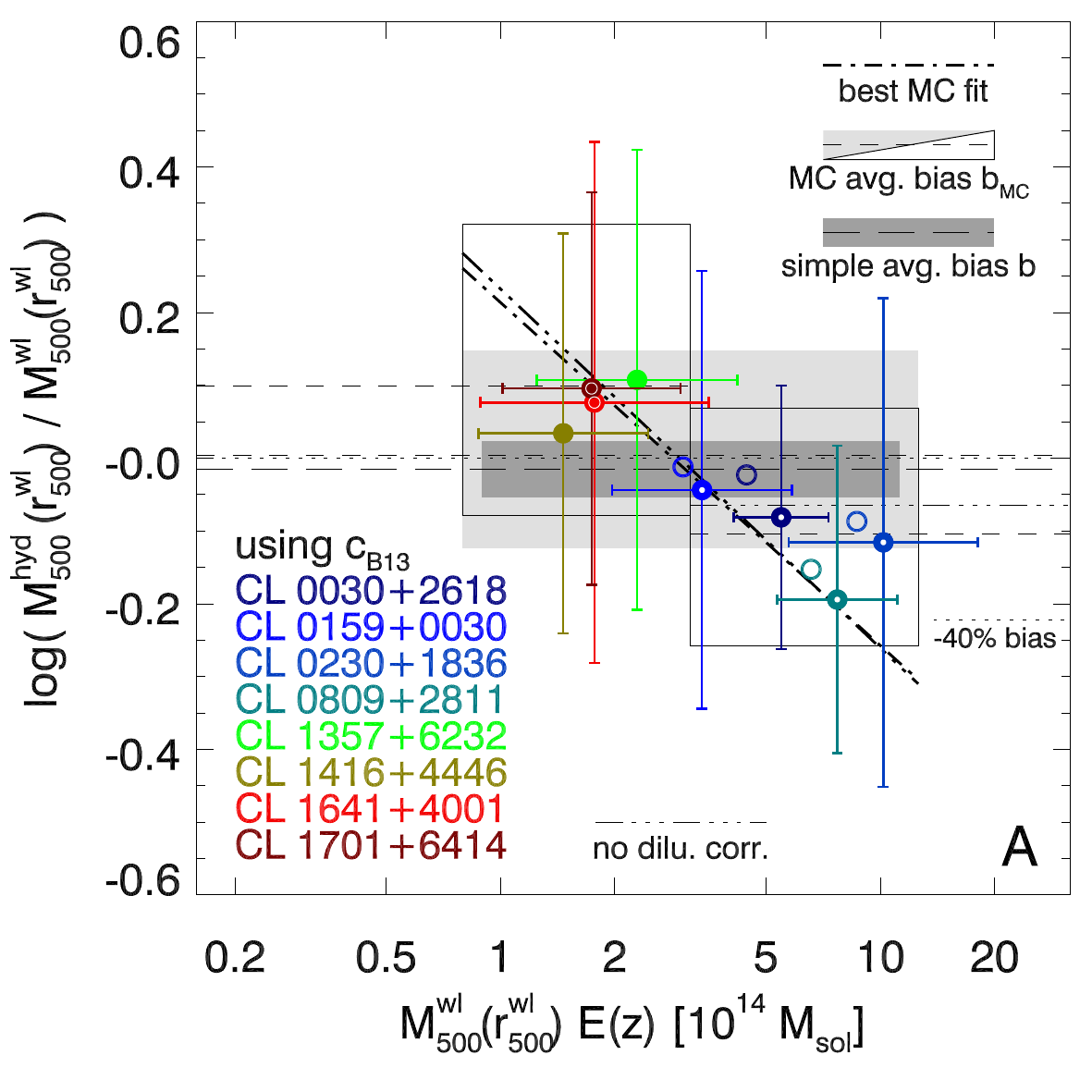}
 \includegraphics[width=8cm]{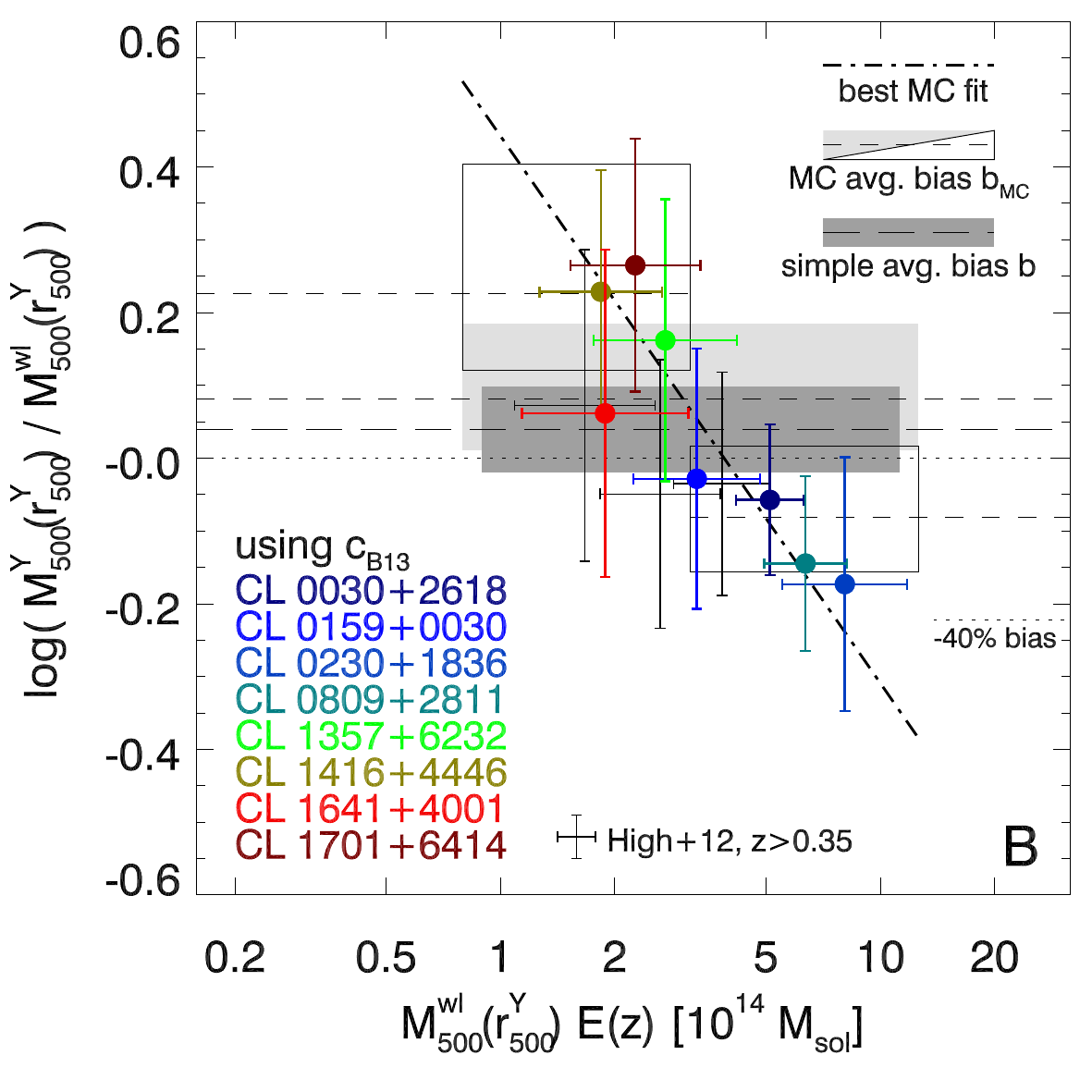}\\
 \end{center}
 \caption{Ratios between X-ray and WL masses as a function of WL mass. 
Panel~A shows $\log{(M^{\mathrm{hyd}}/M^{\mathrm{wl}})}$ within $r_{500}^{\mathrm{wl}}$, 
Panel~B shows $\log{(M^{\mathrm{Y}}/M^{\mathrm{wl}})}$ within $r_{500}^{\mathrm{Y}}$.
WL masses assume the B13 $c$--$M$ relation. We show three tests for a mass bias:
The overall average logarithmic bias $b\!=\!\langle\log M^{\mathrm{X}}\!-\!\log M^{\mathrm{wl}}\rangle$ 
is denoted by a long-dashed line, and its standard error by a dark grey shading.
Short-dashed lines and light grey shading denote the same quantity, but obtained from averaging over
Monte Carlo realisations including the jackknife test. We also show this $b_{\mathrm{MC}}$
for the low-$M^{\mathrm{wl}}$ and high-$M^{\mathrm{wl}}$ clusters separately, with the
$1\sigma$ uncertainties presented as boxes, for sake of clarity. As a visual aid, a dot-dashed line depicts 
the Monte Carlo/jackknife best-fit of $\log{(M^{\mathrm{X}}/M^{\mathrm{wl}})}$ as a function of
$M^{\mathrm{wl}}$. In addition, Panel A also contains this best-fit line (triple-dot-dashed) for the case
without correction for cluster member dilution; the corresponding data points follow the 
Fig.~\ref{fig:mxml} scheme. Indicated by uncertainty bars, Panel B also presents three high-$z$ 
clusters from \citet{2012ApJ...758...68H}.}
  \label{fig:myml}
\end{figure*}
Figures~\ref{fig:myml} and \ref{fig:abiases}, as well as 
Table~\ref{tab:slopes} present the three different X-ray/WL mass bias
estimates for four X-ray mass observables: $M^{\mathrm{hyd}}_{500}$ using $c_{\mathrm{B13}}$
from Sect.~\ref{sec:xl} in Panel A of Fig.~\ref{fig:myml}, $M^{\mathrm{Y}}_{500}$ in Panel B of 
Fig.~\ref{fig:myml}, $M^{\mathrm{T}}_{500}$ in Panel A of Fig.~\ref{fig:abiases}, 
$M^{\mathrm{G}}_{500}$ in Panel B of Fig.~\ref{fig:abiases}. 
The last three are the proxy-based 
\citet{2009ApJ...692.1033V} X-ray mass estimators defined in Sect.~\ref{sec:massdef}.
While Panel A uses $r_{500}^{\mathrm{wl}}$, the other three panel use the respective
$r_{500}^{\mathrm{Y,T,G}}$. 

Long-dashed lines and dark-grey boxes in Fig.~\ref{fig:myml} display $b$ and its error.
Short-dashed lines and light-grey boxes denote $b_{\mathrm{MC}}$.
The intercept $A$ is located at the intersection of the dot-dashed fit and dotted zero lines.
We also show $b_{\mathrm{MC}}$ for the low-mass and high-mass clusters separately, 
splitting at $M^{\mathrm{wl}}E(z)\!=\!M_{\mathrm{piv}}$; the respective
$1\sigma$ ranges are shown as outline boxes.

We observe remarkable agreement between the X-ray/WL mass ratios and bias fits from all four
X-ray observables (which are not fully independent). For each of them, all three bias estimate
agree with each other, and all are consistent with no X-ray/WL mass bias.
We find no evidence for X-ray masses being biased low by $\sim\!\!40$~\% in our cluster sample,
as it has been suggested to explain the \textit{Planck} CMB -- SZ cluster counts discrepancy.
While $b_{\mathrm{MC}}\!\geq\!0.2$ ($\gtrsim\!35$~\% mass bias) lies within the possible range
of the high-$M^{\mathrm{wl}}$ bin, in particular using the gas mass $M^{\mathrm{G}}$,
the overall cluster sample does not support this hypothesis. We point out that $b_{\mathrm{MC}}$
was designed to be both robust against possible effects of large uncertainties and the
small number of clusters in this first batch of \emph{400d} WL clusters.
The larger uncertainties in $b_{\mathrm{MC}}$ compared to $b$ are directly caused by
the jackknife test and the account for possible fit instability in the MC method.

The slopes $B$ quantifying the $M^{\mathrm{wl}}$ dependence of the X-ray/WL mass ratio
are significantly negative in all of our measurements. This directly corresponds to the steep
slope in the mass-mass scaling (Fig.~\ref{fig:mxml}). Predicting cluster masses for very massive
clusters (or low-mass groups) from $B\!=\!\!-0.75$ for $M^{\mathrm{Y}}$ would yield 
$M^{\mathrm{X}}\!=\!M^{\mathrm{wl}}/2$ at $10^{15}\,\mbox{M}_{\sun}/E(z)$ 
(and $M^{\mathrm{X}}\!=\!2.8M^{\mathrm{wl}}$ at $10^{14}\,\mbox{M}_{\sun}/E(z)$). 
Such ratios are at odds with existing measurements. Therefore, we refrain from extrapolating cluster
masses, but interpret the slopes $B$ as indicative of a possibly mass-dependent X-ray/WL mass ratio.
This evidence is more prudently presented as the $\sim\!\!2\sigma$ discrepancy between the 
low-$M^{\mathrm{wl}}$ and high-$M^{\mathrm{wl}}$ mass bins for all three V09a X-ray observables.

\section{A mass-dependent bias?} \label{sec:puzzle}

In this Section, we analyse two unexpected outcomes of our study
in greater detail: the clear correlation between the  
$M^{\mathrm{X}}/M^{\mathrm{wl}}$ measurement of the individual clusters
and their lensing masses, and the unusually small scatter in our scaling relations. 
Results for ancillary scaling relations that we present in Appendix~\ref{sec:anc}
underpin the findings of Fig.~\ref{fig:mxml} and Table~\ref{tab:slopes}.
To begin with, we emphasise that the mass-dependent
bias is not caused by the conflation of a large 
$z$ range; all but one of our clusters inhabit the range $0.39\!\leq\!z\!\leq\!0.53$
across which $E(z)$ varies by $<\!10$~\%, and we accounted for this
variation. As Fig.~\ref{fig:abiases} shows, this also leaves us with little 
constraining power with regard to a $z$-dependent bias, at least until
the \emph{400d} WL survey becomes more complete.

\subsection{Role of $c_{200}$ and departures from NFW profile} \label{sec:c200}

Figure~\ref{fig:mxml} shows that the $M^{\mathrm{wl}}$--$M^{\mathrm{X}}$
scaling relation sensitively depends on the choice
for the cluster concentration parameter $c_{200}$. This translates into
more positive bias estimates for $c_{\mathrm{fit}}$ as compared to $c_{\mathrm{B13}}$
(Table~\ref{tab:slopes}). The difference is caused by the two flat-profile clusters for which
NFW fits yield low masses but do not capture all the large-scale mass distribution,
in particular if $c_{200}$ is determined directly from the data, rather than assuming a 
mass-concentration relation
\citep[cf.\ the discussion of the concentration parameter in Paper~II and][]{2012A&A...546A.106F}.
This induces a bias towards low masses in the $r_{200}\!\!\rightarrow\!\!r_{500}$ conversion.
If these two ``irregular'' clusters (dotted ring symbols in Fig.~\ref{fig:mxml})
are excluded, the ``regular clusters only'' $M^{\mathrm{wl}}$--$M^{\mathrm{X}}$ 
scaling relations (dash-dotted lines) differ for the $c_{\mathrm{fit}}$, but not for the 
$c_{\mathrm{B13}}$ case. Moreover, their mass ratios are consistent with the
other high-$M^{\mathrm{wl}}$ clusters for $c_{\mathrm{B13}}$.
We thus confirm that assuming a mass--concentration relation and marginalising
over $c_{200}$ is advantageous for scaling relations.
We note that \citet{2010ApJ...715..162C} observed a correlation between the scatters in the 
mass--concentration and mass--temperature relations and advocated the inclusion of unrelaxed
clusters in scaling relation studies.

Because NFW profile fits do not capture the complete  projected mass
morphologies of irregular clusters, the assumption of that profile 
for $M^{\mathrm{wl}}_{500}(r_{500}^{\mathrm{wl}})$ (Eq.~\ref{eq:xi}), and 
$M^{\mathrm{wl}}_{500}(r_{500}^{\mathrm{Y}})$, etc.\ (Sect.~\ref{sec:ygt})
could introduce a further bias. 
Aperture-based lensing masses, e.g.\ the $\zeta_{\mathrm{c}}$ statistics 
\citep{1998ApJ...497L..61C} employed by \citet{2012MNRAS.427.1298H} provide an alternative.
However, \citet{2013ApJ...769L..35O} demonstrated by the  stacking of $50$ clusters from
the \textit{Local Cluster Substructure Survey} ($0.15\!<\!z\!<\!0.3$), that the average
weak lensing profile does follow NFW to a high degree, at least at low redshift.
Furthermore, the \citet{2013A&A...550A.129P} finds a trend of $M^{\mathrm{wl}}/M^{\mathrm{hyd}}$
with the ratio of concentration parameters measured from weak lensing and X-rays.

\subsection{Correlation between mass estimates} \label{sec:chisqdisc}

In Table~\ref{tab:slopes}, we quote $\chi^{2}_{\mathrm{red}}$
for the $M^{\mathrm{wl}}$--$M^{\mathrm{X}}$ scaling relations, using the same MC/jackknife
samples as for the bias tests. For the ones involving hydrostatic masses, we measure
$0.5\!\!<\!\!\chi^{2}_{\mathrm{red}}\!\!<\!\!0.6$. We evaluated 
$M^{\mathrm{hyd}}_{500}$ at 
$r_{500}^{\mathrm{wl}}$ in order to measure both estimates within the same physical radius, 
in an ``apples with apples'' comparison. But using the lensing-derived radius
also introduces an unknown amount of correlation, a possible (partial) 
cause of  the measured low $\chi^{2}_{\mathrm{red}}$ values.

We test for the impact of the correlation by measuring both $M^{\mathrm{wl}}$ and $M^{\mathrm{hyd}}$ 
within a fixed physical radius \emph{for all clusters}, and choose 
$r_{\mathrm{fix}}\!=\!800\,\mathrm{kpc}$ as a rough sample average of $r_{500}$.
Surprisingly, we find an only slightly higher $\chi^{2}_{\mathrm{red}}$, still $<\!1$ 
(see Table~\ref{tab:slopes}). 
As before, the bias estimators are consistent with zero. 
Fixed radii of $600\,\mathrm{kpc}$ and $1000\,\mathrm{kpc}$ give similar results
(Table~\ref{tab:slopes} and Fig.~\ref{fig:abiases}). 
with a tentative trend of increasing $\chi^{2}_{\mathrm{red}}$ with smaller radii.
Interestingly, a low $\chi^{2}_{\mathrm{red}}$ is also found for the 
$M^{\mathrm{wl}}$--$M^{\mathrm{T}}$ and $M^{\mathrm{T}}$--$M^{\mathrm{hyd}}$ 
relations (Table~\ref{tab:aslopes}). The latter is expected, because $M^{\mathrm{hyd}}$
are derived from the same $T_{\mathrm{X}}$ and depend sensitively on them.
This all suggests that the small scatter is not driven by 
using $r_{500}^{\mathrm{wl}}$, but by some other intrinsic factor.

If the uncertainties in $M^{\mathrm{wl}}$ were overestimated significantly, 
this would obviously explain the low $\chi^{2}_{\mathrm{red}}$.
However, we do not even include systematic effects here. Moreover, the quoted 
$M^{\mathrm{wl}}$ uncertainties directly come from the NFW modelling of Paper~II
and reflect the $\Delta\chi^{2}$ from their Eq.~(3), given the shear catalogue.
The errors are dominated by the intrinsic
source ellipticity $\sigma_{\varepsilon}$, for which we, after shear calibration,
find values of $\sim\!\!0.3$, consistent with other ground-based WL experiments.
Therefore, despite the allure of our WL errors being overestimated, 
we do not find evidence for this hypothesis in our shear catalogues.
Furthermore, the quoted $M^{\mathrm{wl}}$ uncertainties are consistent with the
aperture mass detection significances we reported in Paper~II.

\subsection{Dilution by cluster members and foregrounds} \label{sec:ndc}

Comparing the complete set of mass-mass scaling relations our data offer (Table~\ref{tab:aslopes}),
we trace the mass-dependence of the bias seen in Fig.~\ref{fig:myml} back to 
the different ranges spanned by the estimates
for $r_{500}$. While the ratio of minimum to maximum is $\approx\!0.75$ for $r_{500}^{\mathrm{Y}}$,
$r_{500}^{\mathrm{T}}$, and $r_{500}^{\mathrm{hyd}}$, the same ratio is $0.57$ for 
$r_{500}^{\mathrm{wl}}$, using the B13 $M$--$c$ relation.
In the following, we discuss the potential influence of several sources of uncertainty in the
WL masses, showing that the dispersion between lowest and highest 
$M^{\mathrm{wl}}_{500}$ is likely an inherent feature rather than a modelling artefact. 

In Paper~II, we discussed the great effort we took in constructing the best-possible homogeneous
analysis from the quite heterogeneous MMT imaging data. Unfortunately, we happen to find higher
$M^{\mathrm{wl}}$ for all clusters with imaging in three bands than for the clusters with
imaging in one band. We emphasise there are no trends with limiting magnitude, seeing, or 
density $n_{\mathrm{KSB}}$ of galaxies with measurable shape (cf.\ Tables~1 and 2 in Paper~II).

In the cases where three-band imaging is available, our WL model includes a correction for the
diluting effect residual cluster member galaxies impose on the shear catalogue. For the other
clusters, no such dilution correction could be applied.
A rough estimate of the fraction of unlensed  galaxies remaining after background selection
suggests that the contamination in single-band catalogues is $\sim\!30$--$50$~\% higher than
with the more sophisticated galaxy-colour based method. 
Therefore, we re-calculate the scaling relations, switching off the dilution correction
(long dashed line and thin ring symbols in Fig.~\ref{fig:mxml}). This lowers the
$r_{500}$ values by $\sim\!\!6$~\% and the masses by $10$--$15$~\%. For 
both the $M^{\mathrm{wl}}_{500}$--$M^{\mathrm{hyd}}_{500}$ and
$M^{\mathrm{wl}}_{500}$--$M^{\mathrm{Y}}_{500}$ relations,
we only observe a slightly smaller difference between $b_{\mathrm{MC}}$ 
for the high- and low-$M^{\mathrm{wl}}$ bins (Tables~\ref{tab:slopes} and \ref{tab:aslopes}), 
not significant given the uncertainty margins.

The dilution of the shear signal by an increased number of galaxies not bearing a shear signal
can also be expressed as a overestimation of the mean lensing depth $\langle\beta\rangle$.
We model a possible lensing depth bias by simultaneously adding the uncertainty 
$\sigma(\langle\beta\rangle)$ for the three-band clusters and subtracting it for the single-band
ones, maximising the leveraging effect on the masses. 
Similar to the previous experiment\footnote{Because an unnoticed higher dilution in the 
catalogue does not imply a bias in the estimation of $\langle\beta\rangle$ from a proxy 
catalogue, the two effects are not likely to add up.} we still observe a mass-dependent bias,
with little change to the default model. 

We further tested alternative choices of cluster centre and fitting range, but do not
observe significant changes to the mass dependent-bias or to $\chi^{2}_{\mathrm{red}}$
(see Appendix~\ref{sec:theta}) .
Although $\langle\beta\rangle$ is calculated for all clusters from the same
catalogues, related systematics would affect the mass normalisation, but not the relative stochastic
uncertainties, which determine $\chi^{2}_{\mathrm{red}}$.
As we consider a drastic overestimation of the purely
statistic uncertainties in the WL modelling being 
unlikely (Sect.~\ref{sec:chisqdisc}), the cause of the low $\chi^{2}_{\mathrm{red}}$ values remains elusive. 
If a WL analysis effect is responsible for one or both anomalies, it has to be 
of a more subtle nature than the choices investigated here.

\subsection{A statistical fluke?} \label{sec:fluke}

We summarise that the $M^{\mathrm{wl}}_{500}$--$M^{\mathrm{hyd}}_{500}$
scaling relations we observe show an unusual lack of scatter and that we find a
$\sim\!2\sigma$ difference between the X-ray/WL mass ratios of our
high- and low-$M^{\mathrm{wl}}$ clusters. The latter 
effect can be traced back to the considerable span in cluster lensing signal,
which is only partly due to 
different background selection procedures and the dilution correction that was only
applied for clusters imaged in three bands.

The question then arises if the mass-dependent bias is 
caused by an unlucky selection of the
$8$ MMT clusters from V09's  wider sample of $36$. 
The $8$ clusters were chosen to be observed first merely because of convenient telescope
scheduling, and appear typical of the larger sample in terms of redshift and X-ray observables.
The MMT clusters trace well the mass range and dispersion spanned by all $36$ clusters in their 
$M^{\mathrm{Y}}_{500}$--$M^{\mathrm{T}}_{500}$ relation. 
We observe the expected vanishing slopes for $\log{(M^{\mathrm{T}}/M^{\mathrm{Y}})}$ 
as a function of $M^{\mathrm{Y}}$, both for the $8$ MMT clusters and for the 
complete sample of $36$ (Table~\ref{tab:aslopes}).

In Table~\ref{tab:slopes}, we observe significant scatter ($\chi^{2}_{\mathrm{red}}\!=\!2.11$) in the 
$M^{\mathrm{wl}}_{500}$--$M^{\mathrm{G}}_{500}$ relation, while
\citet{2010ApJ...721..875O} and M13 reported particularly low scatter in $M^{\mathrm{G}}$, comparing to WL masses.
This large intrinsic scatter seems
to be a feature of the overall \emph{400d} sample: Plotting $M^{\mathrm{G}}$ versus the two other
V09 X-ray masses of all $36$ clusters, we also find $\chi^{2}_{\mathrm{red}}\!>\!2$ (Table~\ref{tab:aslopes}),
as well as significant non-zero logarithmic biases.
While tracing the cause of this observation is beyond the
scope of this article, it deserves further study. Because 
two of the clusters with highest 
$\left|M^{\mathrm{Y,T}}_{500}\!-\!M^{\mathrm{G}}_{500}\right|$ are covered by our MMT subsample, 
we observe a more mass-dependent $M^{\mathrm{G}}/M^{\mathrm{Y,T}}$ ratio than for all $36$.
Overall, however, the MMT subsample is not a very biased selection.

\subsection{Physical causes} \label{sec:physics}

\begin{figure*}
 \includegraphics[width=17cm]{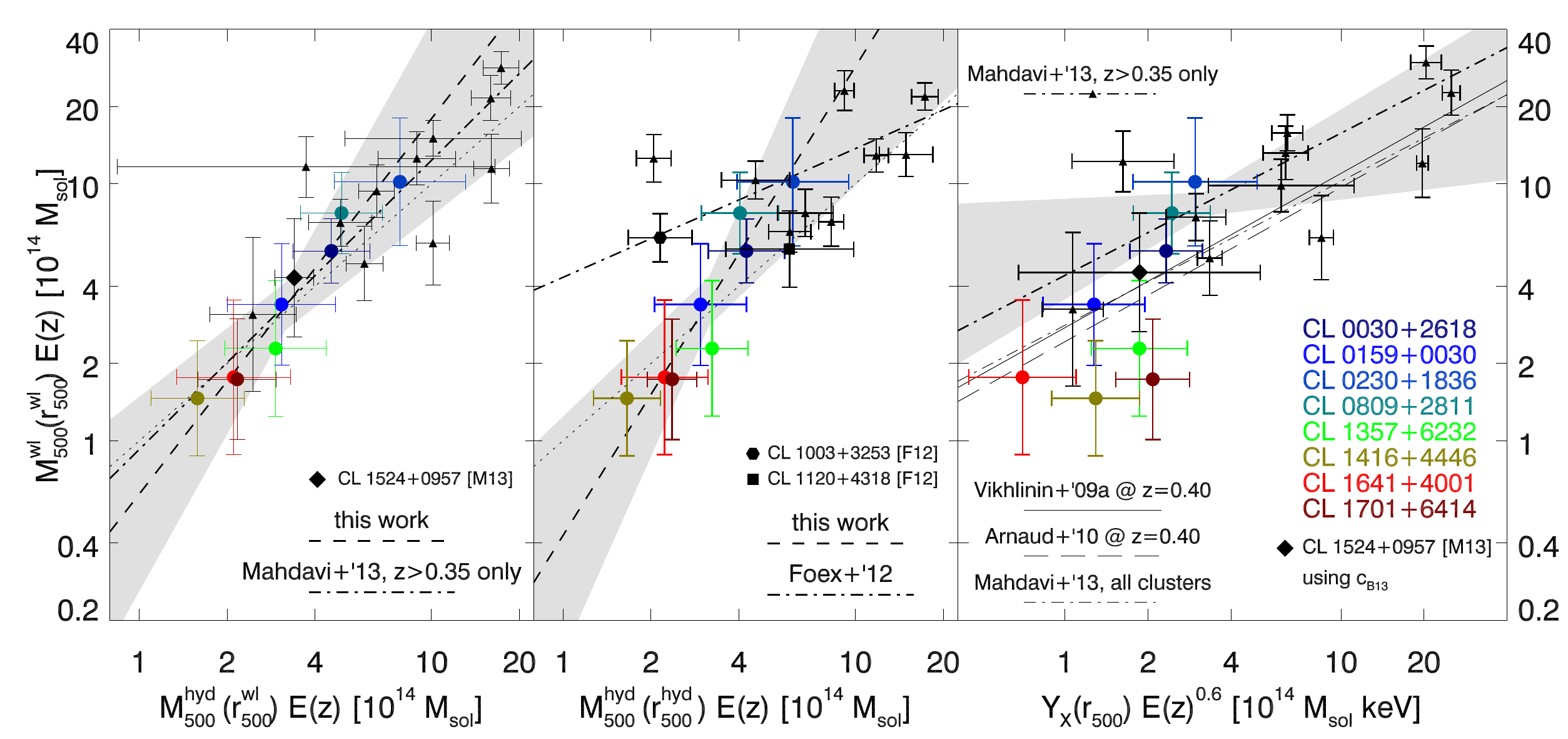}
 \caption{Comparisons with literature data.
 \textit{Left panel:} Black symbols show $z\!>\!0.35$ clusters from 
 \citet{2013ApJ...767..116M}, whose best-fit using Eq.~(\ref{eq:fitexy})
 is shown by the dot-dashed line. The cluster CL\,1524+0957 is indicated by a diamond symbol.
 Coloured symbols and the dashed line show the ``default'' 
 $M^{\mathrm{wl}}_{500}$--$M^{\mathrm{hyd}}_{500}$ relation for $c_{\mathrm{B13}}$ as in the
 lower panel of Fig.~\ref{fig:mxml}. 
 \textit{Middle panel:} The same, but comparing to \citet{2012A&A...546A.106F} 
 (black symbols and dot-dashed line for best fit).
 X-ray masses are measured within $r_{500}^{\mathrm{hyd}}$. 
  CL\,1003+3253 and CL\,1120+4318 are emphasised by special symbols. 
 \textit{Right panel:} Scaling of lensing masses $M^{\mathrm{wl}}_{500}$ with the $Y_{\mathrm{X}}$ proxy. 
  Black symbols show the $z\!>\!0.35$ clusters from M13, to which the thick, dash-dotted line is
  the best fit. Shaded regions indicate  the uncertainties to this fit. The thin, dash-dotted
  line gives the best fit M13 quote for their complete sample, while the thin solid and
  long-dashed lines mark the $M_{500}$--$Y_{\mathrm{X}}$ relations by V09a and 
  \citet{2010A&A...517A..92A}, respectively, for $z\!=\!0.40$.}
 \label{fig:lit}
\end{figure*}
An alternative and likely explanation for the mass-dependent bias we observe could be a high
rate of unrelaxed clusters, especially for our least massive objects. If the departure from
hydrostatic equilibrium were stronger among the low-mass clusters than for the massive ones,
this would manifest in mass ratios similar to our results. Simulations show the offset from
hydrostatic equilibrium to be mass-dependent \citep{2012NJPh...14e5018R},
despite currently being focused on the high-mass regime.
Variability in the non-thermal pressure support with mass \citep{2013A&A...555A..66L} 
may be exacerbated by small number statistics. At high $z$, the 
expected fraction of merging clusters, especially of major mergers, 
increases. Unrelaxed cluster states are known to affect
X-ray observables and, via the NFW fitting, also lensing mass estimates. Indeed, the two
most deviant systems in Fig.~\ref{fig:myml} are CL\,1416+4446 and 
the flat-profile ``shear plateau'' cluster CL\,1641+4001.  
Although the first shows an inconspicuous shear profile, we
suspect it to be part of a possibly interacting supercluster, based on the presence of two
nearby structures at the same redshift, detected in X-ray as well as in our lensing maps 
(Paper~II).
Both these clusters are classified as non-mergers in the recent
\citet{2013arXiv1309.7044N} study, introducing a new substructure estimator 
based on X-ray morphology. However, WL and X-ray methods are sensitive to substructure
on different radial and mass scales, such that this explanation cannot be ruled out.  
We summarise that  the greater dynamical range in WL than in X-ray masses might be 
linked to different sensitivities of the respective methods to substructure and mergers in
the low-mass, high-$z$ cluster population we are probing, but which  is currently still underexplored.

\section{Comparison with previous works} \label{sec:m13disc}

\subsection{The $M^{\mathrm{wl}}_{500}$--$M^{\mathrm{hyd}}_{500}$ relation} \label{sec:hydlit}

\paragraph{Comparison with \citet{2013ApJ...767..116M} results}
Recently, M13  published scaling relations observed between the weak 
lensing and X-ray masses for a sample of $50$ massive clusters, partly based on
the brightest clusters from the \textit{Einstein Observatory} Extended Medium Sensitivity Survey
\citep{1990ApJS...72..567G}. Weak lensing masses for the M13 sample have been obtained from 
CFHT/Megacam imaging \citep{2012MNRAS.427.1298H}, while the X-ray analysis combines 
\textit{XMM-Newton} and \textit{Chandra} data. While the median redshift is $z=0.23$, the
distribution extends to $z\!=\!0.55$, including $12$ clusters at $z\!>\!0.35$. Owing to their
selection, these $12$ clusters lie above the \emph{400d} flux and luminosity cuts, making them
directly comparable to our sample.

The left panel in Fig.~\ref{fig:lit} superimposes the $M^{\mathrm{wl}}_{500}$ and
$M^{\mathrm{hyd}}_{500}$ of the M13 high-$z$ clusters on our results. 
The two samples overlap at the massive ($\gtrsim\!5\!\times\!10^{14}\,\mathrm{M}_{\odot}$)
end, but the \emph{400d} objects probe down to 
$1\!\times\!10^{14}\,\mathrm{M}_{\odot}$ for the first time at this $z$
and for these scaling relations. The slopes of the scaling relations are consistent: 
Using Eq.~(\ref{eq:fitexy}), we measure $B_{\mathrm{M-M}}\!=\!1.13\pm0.20$ for the $12$ M13 objects. 
A joint fit with the \emph{400d} clusters ($B_{\mathrm{M-M}}\!=\!1.46\pm0.57$) yields 
$B_{\mathrm{M-M}}\!\!=\!\!1.15\!\pm\!0.14$ and a low 
$\chi^{2}_{\mathrm{red}}\!=\!0.54$, driven by our data.
We note that the logarithmic bias of $b\!\!=\!\!0.10\!\pm\!0.05$ for the M13 high-$z$ clusters
corresponds to a $(20\!\pm\!10)$~\% mass bias, consistent with both the upper range of the \emph{400d}
results and expectation from the literature \citep[e.g.][]{2010A&A...510A..76L,2012NJPh...14e5018R}. 

Calculating the \citet[H10]{2010arXiv1008.4686H} likelihood which 
\citet{2013ApJ...767..116M} use, we find $B_{\mathrm{H10}}\!=\!1.18^{+0.22}_{-0.20}$ and 
intrinsic scatter $\sigma_{\mathrm{int}}$ consistent with zero, confirming 
our above results. If we, however, repeating our fits from Fig.~\ref{fig:mxml} with the H10
likelihood, we obtain  discrepant results which highlight 
the differences between the various regression algorithms 
(see Sect.~\ref{sec:regress}).\footnote{In fact, regression lines not only depend on the 
likelihood or definition of the best fit, but also on the algorithm used to find its extremum,
and, if applicable, how uncertainties are transferred from the linear to the logarithmic domain.
Thus, our H10 slopes agree with the ones the web-tool provided by M13 yield, but produce 
different uncertainties.}

CL\,1524+0957 at $z\!=\!0.52$ is the only cluster the \emph{400d} and M13 samples share. Denoted
by a black diamond in Fig.~\ref{fig:lit}, its masses from the M13 
lensing and hydrostatic analyses blend in with the MMT \emph{400d} clusters. 
If it were included in the
$M^{\mathrm{wl}}_{500}$--$M^{\mathrm{hyd}}_{500}$ relation, it would 
not significantly alter the best fit, but  we caution that different 
analysis methods have been employed, e.g.\ M13 reporting aperture
lensing masses based on the $\zeta_{\mathrm{c}}$ statistics.

\paragraph{Comparison with \citet{2011ApJ...737...59J} results}
\citet[J11]{2011ApJ...737...59J} studied $14$ very massive and
distant clusters ($0.83\!<\!z\!<\!1.46$) and found their WL and hydrostatic 
masses $M^{\mathrm{wl}}_{200}$ and $M^{\mathrm{hyd}}_{200}$ to agree well,
similar to our results. However, they caution that their $M^{\mathrm{hyd}}_{200}$ 
were obtained by extrapolating a singular isothermal sphere profile.
Because we doubt that the \textit{Chandra}-based \citet{2006ApJ...640..691V} model
reliably describes the ICM out to such large radii, we refrain from deriving
$M^{\mathrm{hyd}}_{200}$.
Nevertheless, we notice that our the J11 samples not only shows similar
$M^{\mathrm{wl}}_{200}$ than our most massive clusters, but also contains  
the only two \emph{400d} clusters exceeding the redshift of CL\,0230+1836, 
CL\,0152$-$1357 at $z\!=\!0.83$ and CL\,1226+3332 at $z\!=\!0.89$. 
Their planned re-analysis will improve the leverage
of our samples at the high-$z$ end. 

\paragraph{Comparison with \citet{2012A&A...546A.106F} results}
In the middle panel of Fig.~\ref{fig:lit}, 
we compare our results to $11$ clusters from the EXCPRES \textit{XMM-Newton} sample, analysed by
\citet[F12]{2012A&A...546A.106F} and located at a similar redshift range 
($0.41\!\leq\!z\!\leq\!0.61$) as the bulk of our sample. Selected to be representative of the
cluster population at $z\!\approx\!0.5$, these objects have been studied with 
\textit{XMM-Newton} in X-rays and CFHT/Megacam in the optical. \citet{2012A&A...546A.106F}
explicitly quote hydrostatic and lensing masses within their respective radii; thus we also show the
$M^{\mathrm{wl}}_{500}(r_{500}^{\mathrm{wl}})$--$M^{\mathrm{hyd}}_{500}(r_{500}^{\mathrm{hyd}})$.
Again, the more massive of our clusters resemble the F12
sources, with the \emph{400d} MMT sample extending towards lower masses. Indeed, F12 study two
clusters which are part of our sample: These, CL\,1002+3253 at $z\!=\!0.42$ and CL\,1120+4318 at
$z\!=\!0.60$ mark their lowest lensing mass objects. At similar $M^{\mathrm{wl}}_{500}$ on
either sides of the best-fit \emph{400d} scaling relation, their inclusion with the
quoted masses would have no immediate effect on its slope, 
but slightly increase its scatter. 

Bearing in mind that Fig.~\ref{fig:lit} (middle panel)  compares quantities measured at
different radii, we notice that the significantly flat best fit regression line to the F12 
cluster masses, showing a larger dispersion in hydrostatic than in WL masses, as opposed to the
\emph{400d} MMT clusters.
The comparisons in Fig.~\ref{fig:lit}  underscore that while being broadly 
compatible with
each other, different studies are shaped by the fine details of their sample selection and
analysis methods. We will conduct a more detailed  comparison between our results and the ones of
\citet{2012A&A...546A.106F} and \citet{2013ApJ...767..116M} once we re-analysed the CFHT/Megacam
of the three overlapping clusters, having already shown the MMT and CFHT Megacams to produce
consistent lensing catalogues (Paper~II).

\subsection{The $M^{\mathrm{wl}}_{500}$--$M^{\mathrm{Y}}_{500}$ relation} \label{sec:yxlit}

The right panel of Fig.~\ref{fig:lit} investigates 
the scaling behaviour of $M^{\mathrm{wl}}_{500}$ with $Y_{\mathrm{X}}$.   
by comparing  the \emph{400d} MMT clusters to the $z\!>\!0.35$ clusters 
from M13.\footnote{Owing to the availability of data, we need to use  different
definitions of $r_{500}$ for the two data sets.} The difference between the two samples is more
pronounced than in the left panel, with only the low-mass end of the M13 sample, 
including CL\,1524+0957, overlapping with our clusters. 
None of the \emph{400d} MMT 
clusters deviates significantly from the $M_{500}$--$Y_{\mathrm{X}}$ relation applied by V09a in
the derivation of the $M^{\mathrm{Y}}_{500}$ masses we used. 
The V09a $M_{500}$--$Y_{\mathrm{X}}$ relation based on \textit{Chandra} data for low-$z$ 
clusters \citep{2006ApJ...640..691V} is in close agreement to the M13 result for their complete
sample, as well as the widely used \citet{2010A&A...517A..92A} $M_{500}$--$Y_{\mathrm{X}}$ 
relation. For the latter as well as V09a we show the version with a slope fixed to the 
self-similar expectation of $B\!=\!3/5$.
The best fit to the M13 $z\!>\!0.35$ essentially yields the same slope as the complete sample
($B\!=\!0.55\pm0.09$ compared to $B_{\mathrm{H10}}\!=\!0.56\pm0.07$, calculated with the
H10 method). The higher normalisation for the high-$z$ subsample can be  likely explained 
as Malmquist bias due to the effective higher mass limit in the M13 sample selection. 
The incompatibility of the least massive MMT clusters with this fit highlights that we sample
lower mass clusters, which, at the same redshift, are likely to have different physical 
properties.

The $Y_{\mathrm{X}}$ proxy is the X-ray equivalent to the integrated pressure signal $Y_{\mathrm{SZ}}$
seen by SZ observatories. Observations confirm a close $Y_{\mathrm{X}}$--$Y_{\mathrm{SZ}}$ 
correlation, with measured departures from the $1$:$1$ slope considered inconclusive 
\citep{2011ApJ...738...48A,2012ApJ...760...67R}. Performing a cursory comparison with SZ
observations, we included in Fig.~\ref{fig:myml}A data for three $z\!>\!0.35$ clusters 
from \citet{2012ApJ...758...68H} (dashed uncertainty bars), taken from their Fig.~6. 
The abscissa values for the \citet{2012ApJ...758...68H} clusters 
(SPT-CL J2022$-$6323, SPT-CL J2030$-$5638, and SPT-CL J2135$-$5726) show masses based on
$Y_{\mathrm{SZ}}$, derived from South Pole Telescope SZ observations \citep{2013ApJ...763..127R}.
The $M^{\mathrm{wl}}$ are derived from observations with the same Megacam instrument we used for
the \emph{400d} clusters, but after its transfer to the Magellan Clay telescope at Las Campanas
Observatory, Chile. In good agreement with zero bias, 
the \citet{2012ApJ...758...68H} clusters are 
consistent with some of the lower mass \emph{400d} clusters. This result suggests that 
the $Y_{\mathrm{X}}$--$Y_{\mathrm{SZ}}$ equivalence might hold once larger samples at high $z$ and low
masses will become available.

\section{Summary and conclusions} \label{sec:conclu}

In this article, we analysed the scaling  relation between WL and X-ray masses for $8$
galaxy clusters drawn from the \emph{400d} sample of X-ray--luminous $0.35\!\leq\!z\!\leq\!0.89$
clusters. WL masses were measured from the \citet{2012A&A...546A..79I} MMT/Megacam
data, and X-ray masses were  based on the V09a \textit{Chandra} analysis. 
We summarise our main results as follows:

\begin{description}
 \item[1.] We probe the WL--X-ray mass scaling relation, in an unexplored region of the parameter 
  space for the first time: the $z\!\sim\!0.4$--$0.5$ redshift range, 
  down to $1\!\times\!10^{14}\,\mathrm{M}_{\odot}$.
 \item[2.] Using several X-ray mass estimates, we find the WL and X-ray masses to be consistent with
  each other. Most of our clusters are compatible with the 
  $M^{\mathrm{X}}\!=\!M^{\mathrm{wl}}$ line. 
 \item[3.] Assuming the $M^{\mathrm{wl}}$ not to be significantly biased, we do not find evidence
  for a systematic underestimation of the X-ray masses by $\sim\!40$~\%, as suggested as a possible
  solution to the discrepancy between the \textit{Planck} CMB constraints on
  $\Omega_{\mathrm{m}}$ and $\sigma_{8}$ (the normalisation of the matter power spectrum) and the
  \textit{Planck} SZ cluster counts \citep{2013arXiv1303.5080P}. 
  While our results favour a small WL--X-ray mass bias, they are consistent with both
  vanishing bias and the $\sim\!20$~\% favoured by studies
  of non-thermal pressure support. 
  \item[4.] For the mass-mass scaling relations involving  $M^{\mathrm{wl}}$, we observe
  a surprisingly low scatter $0.5\!\!<\!\!\chi^{2}_{\mathrm{red}}\!\!<\!\!0.6$, although we use only stochastic
  uncertainties and allow for correlated errors via a Monte Carlo method. Because the errors in
  $M^{\mathrm{wl}}$ are largely determined by the intrinsic WL shape noise $\sigma_{\varepsilon}$, we
  however deem a drastic overestimation unlikely (Sect.~\ref{sec:chisqdisc}). 
  For the scaling relations involving $M_{\mathrm{G}}$, however, we observe 
  a large scatter, contrary to \citet{2010ApJ...721..875O} and M13. 
  \item[5.] Looking in detail, there are intriguing indications for a mass-dependence
  of the WL--X-ray mass ratios of our relatively low-mass $z\!\sim\!0.4$--$0.5$ clusters.
  We observe a mass bias in the low--$M^{\mathrm{wl}}$ mass bin at the $\sim\!2\sigma$ level
  when splitting the sample at $\log{(M_{\mathrm{piv}}/\mathrm{M}_{\odot})}\!=\!14.5$
  This holds for the masses V09a report based on the 
  $Y_{\mathrm{X}}$, $T_{\mathrm{X}}$, and $M_{\mathrm{G}}$ proxies.
\end{description}

We thoroughly investigate possible causes for the mass-dependent bias 
and tight scaling relations. First (Sect.~\ref{sec:c200}), we confirm that 
by using a mass-concentration relation instead of directly fitting
$c_{200}$ from WL, we already significantly reduced the bias due to conversion
from $r_{200}$ to $r_{500}$. We emphasise that, on average,
the NFW shear profile represents a suitable 
fit for the cluster population \citep[cf.][]{2013ApJ...769L..35O}. 
Measuring $M^{\mathrm{hyd}}$ within $r_{500}^{\mathrm{wl}}$
induces correlation between the data points in Fig.~\ref{fig:mxml}. Removing this correlation by
plotting both masses within a fixed physical radius,
we still find small scatter (Sect.~\ref{sec:chisqdisc}). 

We notice that the mass range occupied by the $M^{\mathrm{wl}}$ exceeds the X-ray mass ranges, 
Partially, this higher WL mass range can be explained by  the correction for dilution by member
galaxies, which could be applied only where colour information was available (Paper~II). 
Coincidentally, this is the case for the more massive half of the MMT sample in terms of  
$M^{\mathrm{wl}}$, thus boosting the range of measured WL masses 
(Sect.~\ref{sec:ndc}).  
This result underscores the importance of correcting for the unavoidable 
inhomogeneities in WL data due to the demanding nature of WL observations
\citep[cf.][]{2012arXiv1208.0605A}.
We find no further indications for biases via the WL analysis. 
Furthermore the tight scaling precludes strong redshift effects, and 
we find that our small MMT subsample is largely 
representative of the complete sample of $36$ clusters, judging from the
$M^{\mathrm{Y}}$--$M^{\mathrm{T}}$ relation (Sect.~\ref{sec:fluke}). 
For the $M^{\mathrm{Y}}$--$M^{\mathrm{G}}$ and
$M^{\mathrm{T}}$--$M^{\mathrm{G}}$ relations, significant scatter
($\chi^{2}_{\mathrm{red}}\!>\!2$) is present in the larger sample.
The former relation also shows indications for a significant bias of 
$M^{\mathrm{Y}}\!\approx\!1.15 M^{\mathrm{G}}$. 

Weak lensing and hydrostatic masses for the \emph{400d} MMT clusters are in good agreement with
the $z\!>\!0.35$ part of the \citet{2013ApJ...767..116M} sample and the 
$M^{\mathrm{wl}}_{500}$--$M^{\mathrm{hyd}}_{500}$ relation derived from it
(Sect.~\ref{sec:hydlit}). 
The M13 and \citet{2012A&A...546A.106F} samples include three \emph{400d} clusters with CFHT WL masses. 
These clusters neither point to significantly higher scatter nor to a less mass-dependent bias 
(Fig.~\ref{fig:lit}).
We are planning a re-analysis of the CFHT data, having demonstrated in Paper~II
that lensing catalogues from MMT and CFHT are nicely compatible.
Such reanalysis is going to be helpful to identify more subtle WL analysis
effects potentially responsible for the steep slopes and
tight correlation of WL and X-ray masses.

An alternative explanation are intrinsic differences in the low-mass cluster population. 
That the \emph{400d} MMT sample probes to slightly lower masses ($1\!\times\!10^{14}\,\mathrm{M}_{\odot}$) 
than M13 or F12 becomes especially obvious from the 
$M^{\mathrm{wl}}$--$Y^{\mathrm{X}}$ relation (Fig.~\ref{fig:lit}, Sect.~\ref{sec:yxlit}). 
Because the \emph{400d} sample is  more representative of the 
$z\!\sim\!0.4$--$0.5$ cluster population, it is likely to contain more significant mergers relative
to the cluster mass, 
skewing mass estimates (Sect.~\ref{sec:physics}). Hence, the \emph{400d} survey might be the
first to see the onset of a mass regime in which cluster physics and substructure lead the 
WL--X-ray scaling to deviate from what is known at higher masses. 
Remarkably, Giles et al.\ (in prep.) are finding a different steep slopes
in their low-mass WL--X-ray scaling analysis. Detailed investigations of how their environment
shapes clusters like CL\,1416+4446 might be necessary to improve our understanding
of the cluster population to be seen by future cosmology surveys. 
Analysis systematics might also behave differently at lower masses.
A turn for WL cluster science towards lower mass
objects, e.g.\ through the completion of the \emph{400d} WL sample, will help addressing the
question of evolution in lensing mass scaling relations.

\begin{acknowledgements}
The authors express their thanks to M.\ Arnaud and EXCPRES collaboration (private communication)
for providing the hydrostatic masses of the \citet{2012A&A...546A.106F} clusters.
We further thank A.\ Mahdavi for providing the masses of the \citet{2013ApJ...767..116M}
clusters via their helpful online interface. HI likes to thank M.\ Klein, J.\ Stott, 
and Y.-Y.\ Zhang, and the audiences of his presentations for useful comments.
The authors thank the anonymous referee for their constructive suggestions. 

HI acknowledges support for this work has come from the Deutsche Forschungsgemeinschaft
(DFG) through Transregional Collaborative Research Centre TR 33 as well as 
through the Schwerpunkt Program 1177  and through 
European Research Council grant MIRG-CT-208994.
THR acknowledges support by the DFG through Heisenberg grant RE 1462/5
and grant RE 1462/6.
TE is supported by the DFG through project ER 327/3-1 and by the Transregional
Collaborative Research Centre TR 33 ``The Dark Universe''. 
RM is supported by a Royal Society University Research Fellowship.

We acknowledge the grant of MMT observation time (program 2007B-0046) through NOAO public access. 
MMT time was also provided through support from the F.\ H.\ Levinson Fund 
of the Silicon Valley Community Foundation.
\end{acknowledgements}

\bibliographystyle{aa}
\bibliography{HIsrael400dIII}

\newpage
\appendix
\section{Further scaling relations and tests} \label{sec:anc}

\begin{table*}
\caption{Continuation of Table~\ref{tab:masses}: further measured properties of the \emph{400d} MMT cluster sample.
 All masses are in units of $10^{14}\,\mathrm{M}_{\odot}$, without  applying the $E(z)$ factor. 
 We state only stochastic uncertainties, i.e.\ do not include systematics. 
 By $c_{\mathrm{fit}}$ and $c_{\mathrm{B13}}$
 we denote the choices for the NFW concentration parameter explained in Sect.~\ref{sec:wla}. We refer
 to Sect.~\ref{sec:ndc} for the introduction of the ``varying $\langle\beta\rangle$'' case.}
  \renewcommand{\arraystretch}{1.1}
 \renewcommand\tabcolsep{3.5pt}
 \begin{center}
  \begin{tabular}{c|cccccccc} \hline\hline
 & CL\,0030 & CL\,0159 & CL\,0230 & CL\,0809 & CL\,1357 & CL\,1416 & CL\,1641 & CL\,1701 \\
 & +2618 & +0030 & +1836 & +2811 & +6232 & +4446 & +4001 & +6414 \\ \hline
 $r_{500}^{\mathrm{wl}}(c_{\mathrm{B13}})$, varying $\langle\beta\rangle$ [kpc] & $895_{-90}^{+84}$ & $829_{-175}^{+142}$ & $909_{-191}^{+159}$ & $1075_{-135}^{+122}$ & $710_{-161}^{+136}$ & $650_{-123}^{+104}$ & $687_{-181}^{+136}$ & $668_{-143}^{+117}$ \\
 $M_{500}^{\mathrm{wl}}(r_{500}^{\mathrm{wl}})$, using $c_{\mathrm{B13}}$, varying $\langle\beta\rangle$ & $3.70_{-1.01}^{+1.14}$ & $2.58_{-1.32}^{+1.57}$ & $5.54_{-2.81}^{+3.45}$ & $5.70_{-1.89}^{+2.17}$ & $1.91_{-1.03}^{+1.32}$ & $1.26_{-0.59}^{+0.71}$ & $1.60_{-0.96}^{+1.15}$ & $1.45_{-0.75}^{+0.90}$ \\
$M^{\mathrm{wl}}(r_{\mathrm{fix}}\!=\!800\,\mbox{kpc})$ & $3.57_{-0.76}^{+0.73}$ & $2.61_{-1.09}^{+0.91}$ & $5.33_{-2.15}^{+1.89}$ & $4.37_{-1.13}^{+1.10}$ & $1.99_{-0.94}^{+0.79}$ & $1.46_{-0.57}^{+0.51}$ & $1.66_{-0.97}^{+0.73}$ & $1.65_{-0.69}^{+0.59}$ \\
 $M_{500}^{\mathrm{wl}}(r_{500}^{\mathrm{Y}})$, using $c_{\mathrm{fit}}$ & $3.71_{-0.76}^{+0.87}$ & $3.78_{-1.55}^{+1.27}$ & $5.30_{-2.15}^{+2.55}$ & $4.08_{-1.21}^{+1.91}$ & $2.08_{-0.87}^{+0.84}$ & $1.52_{-0.63}^{+0.56}$ & $1.05_{-0.27}^{+0.35}$ & $1.42_{-0.49}^{+0.55}$ \\
 $M_{500}^{\mathrm{wl}}(r_{500}^{\mathrm{Y}})$, using $c_{\mathrm{fit}}$, no dilu.\ corr.\ & $3.28_{-0.65}^{+0.72}$ & $3.43_{-1.33}^{+1.13}$ & $4.71_{-1.89}^{+2.21}$ & $3.65_{-1.04}^{+1.53}$ & -- & -- & -- & -- \\
 $M_{500}^{\mathrm{wl}}(r_{500}^{\mathrm{Y}})$, using $c_{\mathrm{B13}}$, no dilu.\ corr.\ & $3.41_{-0.73}^{+0.66}$ & $2.48_{-1.05}^{+0.86}$ & $4.67_{-1.87}^{+1.68}$ & $4.70_{-1.17}^{+1.12}$ & -- & -- & -- & -- \\
 $M_{500}^{\mathrm{hyd}}(r_{500}^{\mathrm{wl}})$, using $c_{\mathrm{B13}}$, varying $\langle\beta\rangle$ & $3.32_{-0.96}^{+1.07}$ & $2.44_{-0.99}^{+1.11}$ & $4.54_{-2.12}^{+2.57}$ & $3.86_{-1.17}^{+1.31}$ & $2.30_{-0.88}^{+0.96}$ & $1.31_{-0.45}^{+0.52}$ & $1.70_{-0.73}^{+0.73}$ & $1.82_{-0.56}^{+0.52}$ \\
$M^{\mathrm{hyd}}(r_{\mathrm{fix}}\!=\!800\,\mbox{kpc})$ & $3.06\pm0.61$ & $2.39\pm0.54$ & $3.84\pm0.71$ & $2.89\pm0.51$ & $2.51\pm0.38$ & $1.62\pm0.23$ & $1.88\pm0.35$ & $2.01\pm0.21$ \\ 
 $M_{500}^{\mathrm{hyd}}(r_{500}^{\mathrm{Y}})$ & $3.29_{-0.88}^{+0.97}$ & $2.44_{-0.77}^{+0.85}$ & $3.67_{-1.14}^{+1.34}$ & $3.41_{-0.86}^{+0.94}$ & $2.55_{-0.58}^{+0.60}$ & $1.66_{-0.42}^{+0.45}$ & $1.73_{-0.54}^{+0.56}$ & $2.15_{-0.37}^{+0.38}$ \\
 $M_{500}^{\mathrm{hyd}}(r_{500}^{\mathrm{T}})$ & $3.52_{-0.88}^{+0.95}$ & $2.48_{-0.76}^{+0.84}$ & $3.73_{-1.14}^{+1.34}$ & $3.16_{-0.89}^{+0.98}$ & $2.52_{-0.59}^{+0.61}$ & $1.47_{-0.46}^{+0.52}$ & $1.74_{-0.54}^{+0.56}$ & $2.04_{-0.40}^{+0.41}$ \\
 $M_{500}^{\mathrm{hyd}}(r_{500}^{\mathrm{G}})$ & $2.83_{-0.93}^{+1.05}$ & $2.26_{-0.79}^{+0.90}$ & $3.21_{-1.13}^{+1.38}$ & $3.49_{-0.86}^{+0.93}$ & $2.42_{-0.62}^{+0.64}$ & $1.77_{-0.40}^{+0.42}$ & $1.64_{-0.60}^{+0.61}$ & $2.13_{-0.37}^{+0.39}$ \\
 \hline\hline
 \end{tabular} 
  \label{tab:amasses}
 \end{center}
\end{table*}
\begin{table*}
 \caption{Continuation of Table~\ref{tab:slopes}: 
We estimate a possible bias between masses $\xi$ and $\eta$ by three estimators: 
First, we fit to $(\log \xi -\log \eta)$ as a function of $\eta$, yielding an intercept $A$ at pivot
$\log{\left(M_{\mathrm{piv}}/\mathrm{M}_{\odot}\right)}\!=\!14.5$ and slope $B$ 
from the Monte Carlo/jackknife analysis. Second, we compute the logarithmic bias 
$b_{\mathrm{MC}}\!=\!\langle\log{\xi}\!-\!\log{\eta}\rangle_{\mathrm{MC}}$, averaged over the same realisations.
Uncertainties for the MC results are given by $1\sigma$ ensemble dispersions.
In parentheses next to $b_{\mathrm{MC}}$, we show its value for the low-$M^{\mathrm{wl}}$ 
and high-$M^{\mathrm{wl}}$ clusters.
Third, we quote the logarithmic bias $b\!=\!\langle\log{\xi}\!-\!\log{\eta}\rangle$ obtained directly
from the input masses, along with its standard error.
Finally, we give the $\chi^{2}_{\mathrm{red}}$ for the mass-mass scaling, obtained from the MC method. 
The ``default'' model denotes WL and hydrostatic masses as described in Sect.~\ref{sec:obsdat}.}
 \renewcommand{\arraystretch}{1.1}
 \renewcommand\tabcolsep{3pt}
 \begin{center}
  \begin{tabular}{ccc|ccccc|c}\hline\hline
Scaling Relation & Model & $c_{\mathrm{NFW}}$ & Slope $B$ & Intercept $A$ & $b_{\mathrm{MC}}$ from Monte Carlo & $b\!=\!\langle\log{\xi}\!-\!\log{\eta}\rangle$ &  $\chi^{2}_{\mathrm{red,M-M}}$ &Section \\ \hline
    $M^{\mathrm{wl}}_{500}(r_{500}^{\mathrm{wl}})$--$M^{\mathrm{hyd}}_{500}(r_{500}^{\mathrm{wl}})$ & 
     varying $\langle\beta\rangle$ & $c_{\mathrm{fit}}$ & $-0.54_{-0.21}^{+0.22}$ & $0.00_{-0.08}^{+0.07}$ & $0.08_{-0.13}^{+0.14}$ ($0.23_{-0.18}^{+0.20}$; $-0.07_{-0.15}^{+0.16}$) &  $0.07\pm0.07$ &$0.60$ & \ref{sec:ndc} \\ 
    & varying $\langle\beta\rangle$ & $c_{\mathrm{B13}}$ & $-0.50_{-0.30}^{+0.30}$ & $-0.02\pm0.08$ & $0.01_{-0.13}^{+0.15}$ ($0.08_{-0.18}^{+0.22}$; $-0.07_{-0.15}^{+0.17}$) & $-0.01\pm0.03$ & $0.50$ & \ref{sec:ndc} \\
    & $S$-peak centred & $c_{\mathrm{fit}}$ & $-0.48_{-0.23}^{+0.24}$ & $-0.02\pm0.07$ & $-0.03_{-0.11}^{+0.13}$ ($0.08_{-0.16}^{+0.19}$; $-0.14_{-0.14}^{+0.15}$) & $-0.04\pm0.04$ &$0.59$ & \ref{sec:theta} \\ 
    & $S$-peak centred & $c_{\mathrm{B13}}$ & $-0.47_{-0.25}^{+0.26}$ & $-0.03\pm0.07$ & $-0.05_{-0.10}^{+0.12}$ ($0.01_{-0.14}^{+0.16}$; $-0.12_{-0.14}^{+0.15}$) & $-0.07\pm0.03$ & $0.60$ & \ref{sec:theta} \\ \hline
 $M^{\mathrm{wl}}_{500}(r_{500}^{\mathrm{Y}})$--$M^{\mathrm{Y}}_{500}(r_{500}^{\mathrm{Y}})$ & 
    default & $c_{\mathrm{fit}}$ & $-0.73_{-0.14}^{+0.13}$ & $0.05\pm0.04$ & $0.08_{-0.06}^{+0.08}$ ($0.26_{-0.08}^{+0.11}$; $-0.10_{-0.08}^{+0.10}$) & $0.06\pm0.07$ & $1.67$& \ref{sec:ygt} \\
    & no dilu.\ corr.\ & $c_{\mathrm{fit}}$ & $-0.71_{-0.15}^{+0.14}$ & $0.06\pm0.04$ & $0.11_{-0.06}^{+0.08}$ ($0.26_{-0.08}^{+0.11}$; $-0.05_{-0.08}^{+0.10}$) & $0.09\pm0.06$ & $1.51$ & \ref{sec:ndc} \\
    & no dilu.\ corr.\ & $c_{\mathrm{B13}}$ & $-0.73_{-0.15}^{+0.13}$ & $0.07\pm0.03$ & $0.10_{-0.07}^{+0.10}$ ($0.23_{-0.11}^{+0.18}$; $-0.04_{-0.07}^{+0.10}$) & $0.06\pm0.05$ & $1.08$ & \ref{sec:ndc} \\ \hline
 $M^{\mathrm{wl}}_{500}(r_{500}^{\mathrm{Y}})$--$M^{\mathrm{hyd}}_{500}(r_{500}^{\mathrm{Y}})$ &
     default & $c_{\mathrm{fit}}$ & $-0.63_{-0.20}^{+0.20}$ & $0.01\pm0.05$  & $0.02_{-0.08}^{+0.09}$ ($0.14_{-0.10}^{+0.12}$; $-0.12_{-0.11}^{+0.12}$) & $0.01\pm0.05$ &$0.77$ & \ref{sec:ygt} \\
 $M^{\mathrm{wl}}_{500}(r_{500}^{\mathrm{Y}})$--$M^{\mathrm{hyd}}_{500}(r_{500}^{\mathrm{Y}})$ &
     default & $c_{\mathrm{B13}}$ & $-0.56_{-0.23}^{+0.23}$ & $0.00\pm0.05$ & $0.02_{-0.09}^{+0.11}$ ($0.12_{-0.12}^{+0.18}$; $-0.10_{-0.10}^{+0.12}$) & $-0.02\pm0.04$ & $0.65$ & \ref{sec:ygt} \\ 
$M^{\mathrm{wl}}_{500}(r_{500}^{\mathrm{T}})$--$M^{\mathrm{hyd}}_{500}(r_{500}^{\mathrm{T}})$ &
      default & $c_{\mathrm{B13}}$ & $-0.53_{-0.22}^{+0.23}$ & $0.04\pm0.05$ & $0.03_{-0.05}^{+0.06}$ ($0.09_{-0.06}^{+0.07}$; $-0.05_{-0.08}^{+0.09}$)& $-0.02\pm0.04$ & $0.66$ & \ref{sec:ygt}\\ 
 $M^{\mathrm{wl}}_{500}(r_{500}^{\mathrm{G}})$--$M^{\mathrm{hyd}}_{500}(r_{500}^{\mathrm{G}})$ &
      default & $c_{\mathrm{B13}}$ & $-0.61_{-0.25}^{+0.24}$ & $-0.01\pm0.06$ & $0.01_{-0.09}^{+0.11}$ ($0.12_{-0.12}^{+0.18}$; $-0.10_{-0.12}^{+0.12}$) & $-0.02\pm0.04$ & $0.62$ & \ref{sec:ygt}\\ \hline
 $M^{\mathrm{hyd}}_{500}(r_{500}^{\mathrm{Y}})$--$M^{\mathrm{Y}}_{500}(r_{500}^{\mathrm{Y}})$ &
     default & -- & $-0.76_{-0.29}^{+0.21}$ & $0.07\pm0.03$ & $0.07\pm0.05$ & $0.06\pm0.03$ &$1.29$ & \ref{sec:ygt} \\
 $M^{\mathrm{hyd}}_{500}(r_{500}^{\mathrm{T}})$--$M^{\mathrm{T}}_{500}(r_{500}^{\mathrm{T}})$ &
     default &  -- & $-0.53_{-0.39}^{+0.37}$ & $0.04\pm0.05$ & $0.04\pm0.07$  & $0.04\pm0.02$ & $0.79$ & \ref{sec:ygt} \\ 
 $M^{\mathrm{hyd}}_{500}(r_{500}^{\mathrm{G}})$--$M^{\mathrm{G}}_{500}(r_{500}^{\mathrm{G}})$ &
      default & -- & $-1.20_{-0.62}^{+0.36}$  & $-0.05_{-0.04}^{+0.03}$ & $0.03_{-0.05}^{+0.06}$  & $0.01\pm0.05$ & $1.85$ & \ref{sec:ygt}\\ \hline
 $M^{\mathrm{Y}}_{500}(r_{500}^{\mathrm{Y}})$--$M^{\mathrm{T}}_{500}(r_{500}^{\mathrm{T}})$ & default (MMT8) & 
 -- & $-0.07_{-0.51}^{+0.65}$ &  $-0.03_{-0.07}^{+0.06}$ &  $-0.04\pm0.05$ & $-0.03\pm0.03$ & $1.27$ & \ref{sec:fluke} \\
 & all 36 & -- & $-0.05_{-0.12}^{+0.13}$ & $0.00\pm0.02$ & $-0.03\pm0.02$ & $-0.02\pm0.01$ & $1.32$ & \ref{sec:fluke} \\ \hline
  $M^{\mathrm{Y}}_{500}(r_{500}^{\mathrm{Y}})$--$M^{\mathrm{G}}_{500}(r_{500}^{\mathrm{G}})$ & default (MMT8) &
 -- &  $-0.39_{-0.55}^{+0.47}$ & $-0.02_{-0.04}^{+0.05}$ & $-0.07\pm0.02$ & $-0.07\pm0.04$ & $3.43$ & \ref{sec:fluke}\\
 & all 36 & -- & $-0.05_{-0.06}^{+0.06}$ & $-0.05\pm0.01$ & $-0.07\pm0.01$ & $-0.07\pm0.01$ & $2.32$ & \ref{sec:fluke}\\ \hline
 $M^{\mathrm{T}}_{500}(r_{500}^{\mathrm{T}})$--$M^{\mathrm{G}}_{500}(r_{500}^{\mathrm{G}})$ & default (MMT8) & 
 -- & $-1.24_{-0.41}^{+0.28}$  & $0.05\pm0.03$ & $-0.03_{-0.04}^{+0.05}$ & $-0.04\pm0.06$ & $2.81$ & \ref{sec:fluke}\\
 & all 36 & -- & $-0.55_{-0.14}^{+0.10}$ & $0.02\pm0.02$ & $-0.04\pm0.02$ & $-0.05\pm0.02$ & $2.58$ & \ref{sec:fluke}\\
   \hline\hline \label{tab:aslopes}
  \end{tabular}
 \end{center}
\end{table*}
\begin{figure}
 \begin{center}
  \includegraphics[width=8cm]{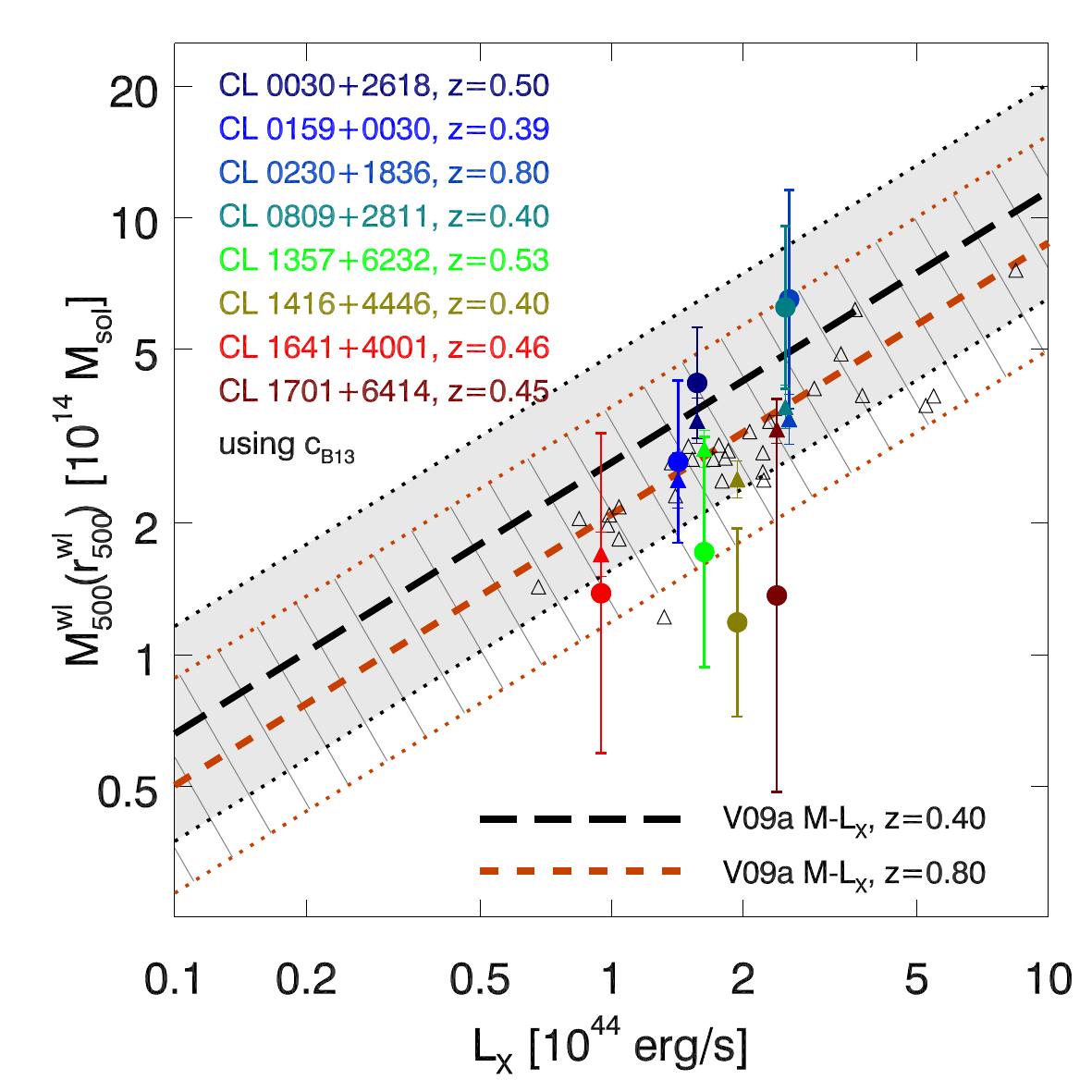}
 \end{center}
 \caption{Lensing mass -- X-ray luminosity relation.
  The $M$--$L_{\mathrm{X}}$ relation is shown, for both 
  $M^{\mathrm{wl}}_{500}(r_{500}^{\mathrm{wl}})$ (filled circles) and
  $M^{\mathrm{Y}}_{500}(r_{500}^{\mathrm{Y}})$ (small triangles).
  Open triangles represent the sample clusters for which MMT lensing masses are not available. 
  The V09a $M$-$L_{\mathrm{X}}$ relation at $z\!=\!0.40$ ($z\!=\!0.80$) is 
  denoted by a long-dashed black (short-dashed red) line. 
  Shaded (hatched) areas show the respective $1\sigma$ 
  intrinsic scatter ranges.}
\label{fig:lxml}
\end{figure}
\begin{figure*}
 \begin{center}
 \includegraphics[width=8cm]{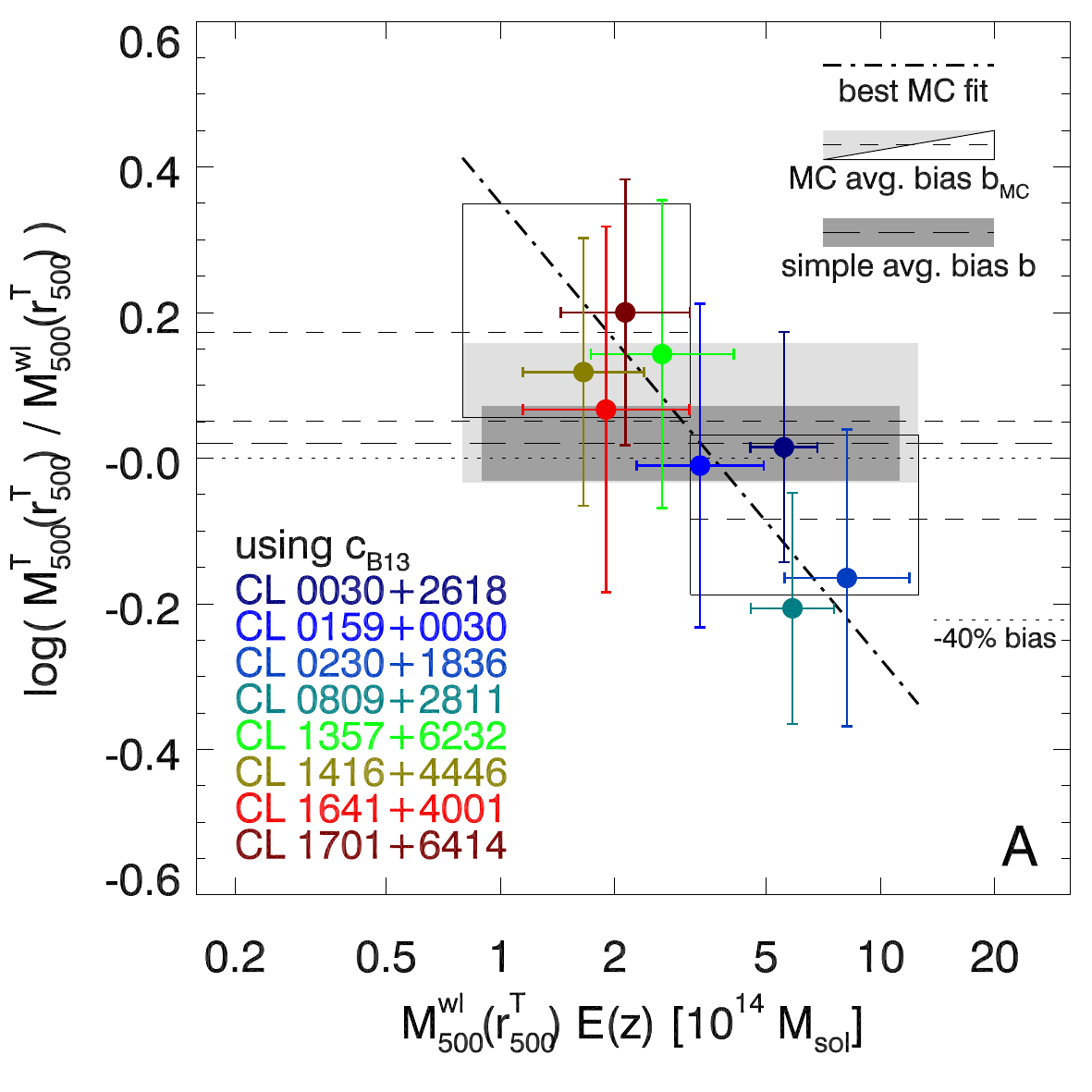}
 \includegraphics[width=8cm]{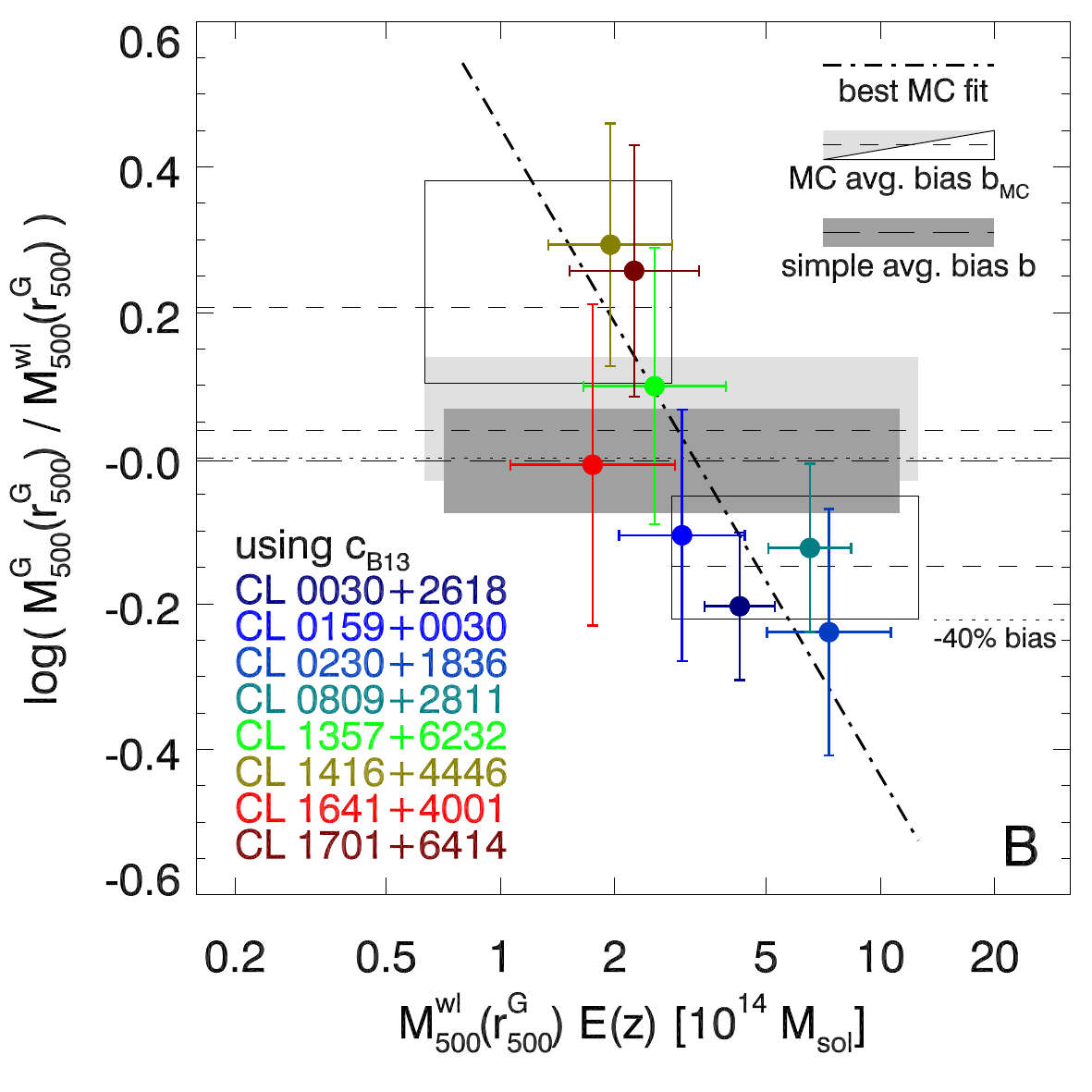}\\
 \includegraphics[width=8cm]{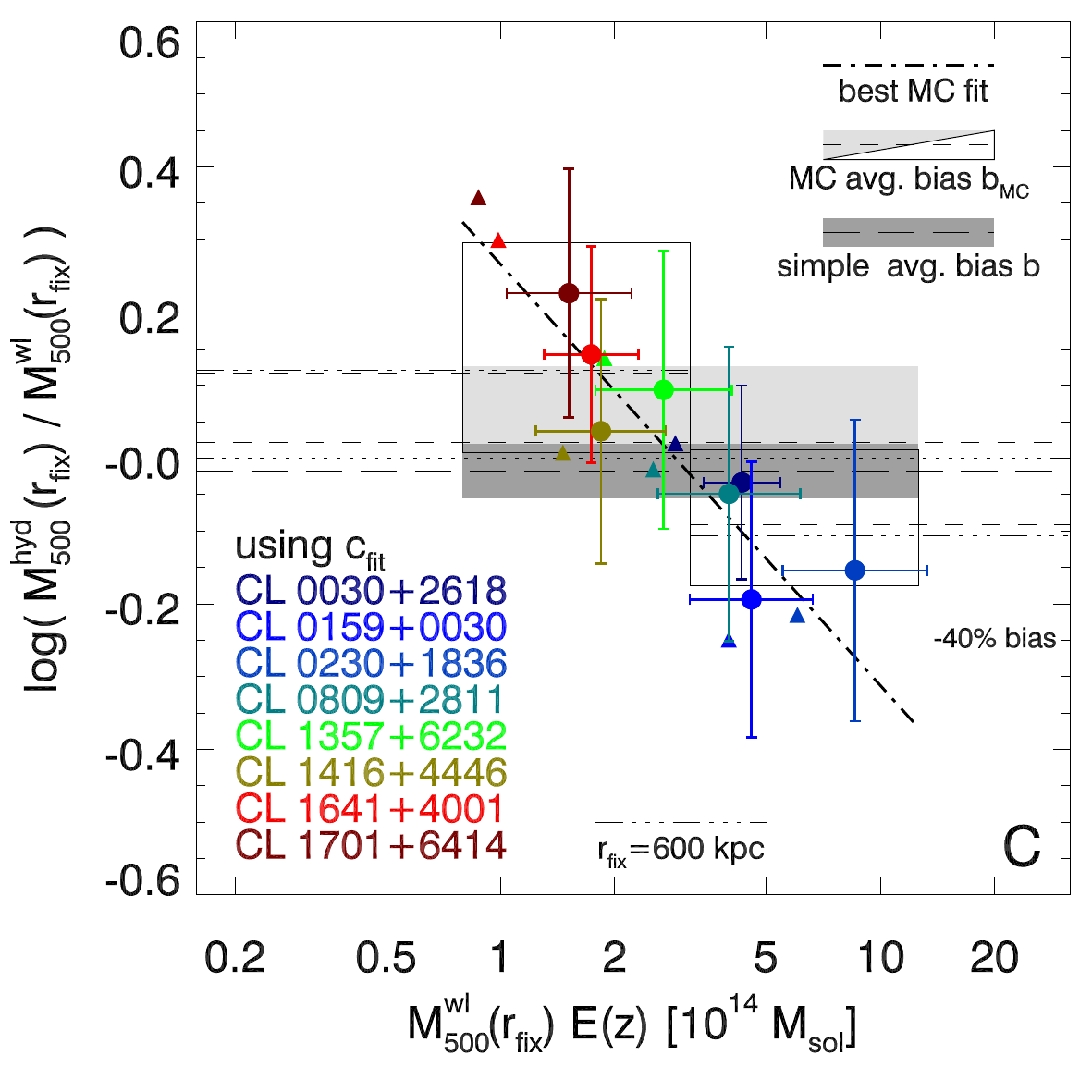}
 \includegraphics[width=8cm]{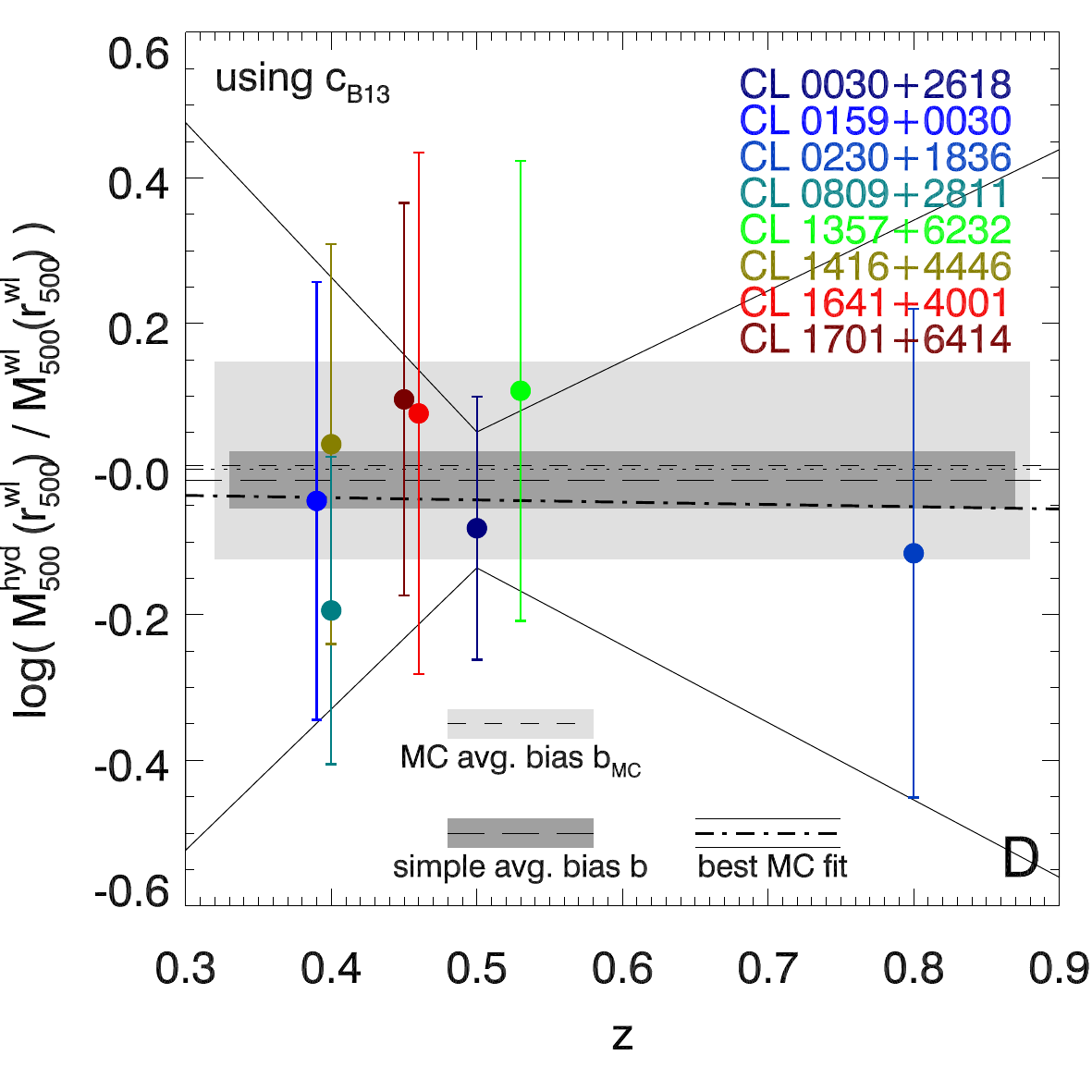}
 \end{center}
 \caption{Continuation of Fig.~\ref{fig:myml}. 
 Panel~A shows $\log{(M^{\mathrm{T}}/M^{\mathrm{wl}})}$ within $r_{500}^{\mathrm{T}}$, 
 Panel~B shows $\log{(M^{\mathrm{G}}/M^{\mathrm{wl}})}$ within $r_{500}^{\mathrm{G}}$.
 Like Panel~A of Fig.~\ref{fig:myml}, Panel~C presents $\log{(M^{\mathrm{hyd}}/M^{\mathrm{wl}})}$, 
 but showing both WL masses measured at a fixed physical radius $r_{\mathrm{fix}}$. Filled dots and  
 dot-dashed lines correspond to $r_{\mathrm{fix}}\!=\!800\,\mbox{kpc}$, 
 while triangles and triple-dot-dashed lines denote $r_{\mathrm{fix}}\!=\!600\,\mbox{kpc}$.
 Uncertainties for the $600\,\mbox{kpc}$ case were omitted for clarity.
 Panel~D shows $\log{(M^{\mathrm{hyd}}/M^{\mathrm{wl}})}$ from Fig.~\ref{fig:myml}
 as a function of redshift. Thin
 solid lines indicating the $1\sigma$ uncertainty range of the best-fit Monte Carlo/jackknife
 regression line (dot-dashed).}
\label{fig:abiases}
\end{figure*}

\subsection{The $L_{\mathrm{X}}$--$M$ relation} \label{sec:lxm}

To better assess the consistency of our weak lensing masses with the \citet{2009ApJ...692.1033V}
results, we compare them to the $L_{\mathrm{X}}$-$M_{\mathrm{Y}}$--relation derived by V09a
using the $M^{\mathrm{Y}}_{500}$ masses \emph{of their low-$z$ cluster sample}.
Figure~\ref{fig:lxml} inverts this relation by showing the
$M^{\mathrm{wl}}_{500}(r_{500}^{\mathrm{wl}})$ masses as a function  of the 
$0.5$--$2.0\,\mbox{keV}$ \textit{Chandra} luminosities measured by V09a. 
Statistical uncertainties in the \textit{Chandra} fluxes and, 
hence, luminosities are negligible for our purposes. 
We calculate the expected $68$~\% confidence ranges in mass for a
given luminosity by inverting the scatter in $L_{\mathrm{X}}$ 
at a fixed $M^{\mathrm{Y}}$ as given in Eq.~(22) of V09a.
For two fiducial redshifts, $z\!=\!0.40$ and $z\!=\!0.80$, spanning the 
unevenly populated redshift range of the eight clusters, the $M$--$L_{\mathrm{X}}$ relations and 
their expected scatter are shown in Fig.~\ref{fig:lxml}.
Small filled triangles in Fig.~\ref{fig:lxml} show the $M^{\mathrm{Y}}_{500}$ masses from 
which V09a derived the $L_{\mathrm{X}}$--$M$ relation. Our $8$ MMT clusters 
are nicely tracing the distribution of the  overall sample of $36$ 
clusters (open triangles).

As an important step in the calculation of the mass function, 
these authors show that  their procedure
is able to correct for the Malmquist bias even in the presence of evolution in the 
$L_{\mathrm{X}}$-$M$ relation, which they include in the model.
We emphasise that the Malmquist bias correction -- which is not included here -- 
applied by V09a moves the clusters upwards in Fig.~\ref{fig:lxml}, such that
the sample agrees  with the best-fit
from the low-$z$ sample, as Fig.~12 in V09a  demonstrates. 

As already seen in Fig.~\ref{fig:myml}, the $M^{\mathrm{wl}}$
(large symbols in Fig.~\ref{fig:lxml}) and $M^{\mathrm{Y}}$ agree well. 
Thus we can conclude that the WL masses are consistent with the
expectations from their $L_{\mathrm{X}}$.
Finally, we remark that the  higher X-ray 
luminosities for the some of the same clusters reported by
\citet{2012MNRAS.421.1583M} in their study of the $L_{\mathrm{X}}$--$T_{\mathrm{X}}$ relation
are not in disagreement with V09a, as \citet{2012MNRAS.421.1583M} used bolometric luminosities.

\subsection{Redshift scaling and cross-scaling of X-ray masses}

Here we show further results mentioned in the main body of the article.
Figure~\ref{fig:abiases} shows two examples of the X-ray/WL mass ratio as a function of redshift.
Owing to the inhomegenous redshift coverage of our clusters, we cannot constrain a redshift
evolution. All of our bias estimates are consistent with zero bias.

Table~\ref{tab:aslopes} shows the fit results and bias estimates for various tests we performed
modifying our default model, as well as for ancillary scaling relations. In particular, we probe
the scaling behaviour of hydrostatic masses against the V09a estimates, for which we find
a $M^{\mathrm{Y}}/M^{\mathrm{hyd}}$ tentatively biased high by $\sim\!15$~\%,
while $M^{\mathrm{T}}$ and $M^{\mathrm{G}}$ do not show similar biases.

\subsection{Choice of centre and fitting range} \label{sec:theta}

Weak lensing masses obtained from profile fitting have been shown to be sensitive to
the choice of the fitting range 
\citep{2011ApJ...740...25B,2011MNRAS.412.2095H,2011MNRAS.414.1851O}. 
Taking these results into
account, we fitted the WL masses within a fixed physical mass range. Varying the
fitting range by using  $r_{\mathrm{min}}\!=\!0$ instead of $0.2\,\mbox{Mpc}$
in one  and $r_{\mathrm{max}}\!=\!4.0\,\mbox{Mpc}$  instead of 
$5.0\,\mbox{Mpc}$) in another test, we find no evidence for  a crucial
influence on our results.

Both simulations and observations establish 
\citep[e.g.][]{2012MNRAS.419.3547D,2012ApJ...757....2G} that WL masses using lensing cluster 
centres are biased high due to random noise with respect to those based on independently 
obtained cluster centres, e.g.\ the \textit{Rosat} centres we employ. The fact that the
$M^{\mathrm{wl}}_{500}$--$M^{\mathrm{hyd}}_{500}$ 
relation gives slightly milder difference between $b_{\mathrm{MC}}$ 
for the high- and low-$M^{\mathrm{wl}}$ bins 
when the peak of the $S$-statistics is assumed as the cluster 
centre (Table~\ref{tab:aslopes})  
can be explained by the larger relative $M^{\mathrm{wl}}$
``boost'' for clusters with larger offset between X-ray and lensing peaks. This affects the
flat-profile clusters (Sect.~\ref{sec:c200}) 
in particular, translating into a greater effect for the $c_{\mathrm{fit}}$
case than for $c_{\mathrm{B13}}$-based masses.  
We find that WL cluster centres only slightly alleviate the observed mass-dependence.

\end{document}